\documentclass[journal]{IEEEtran}
\usepackage{amsmath,amsfonts}
\usepackage{algorithmic}
\usepackage{array}
\usepackage[caption=false,font=normalsize,labelfont=sf,textfont=sf]{subfig}
\usepackage{textcomp}
\usepackage{stfloats}
\usepackage{url}
\usepackage{verbatim}
\usepackage{graphicx}
\def\BibTeX{{\rm B\kern-.05em{\sc i\kern-.025em b}\kern-.08em
    T\kern-.1667em\lower.7ex\hbox{E}\kern-.125emX}}
\usepackage{balance}
\usepackage{enumitem}

\usepackage{booktabs}
\usepackage{multirow}
\usepackage{makecell}
\usepackage{xcolor}

\setlist[itemize]{left=0pt, label=\textbullet, labelsep=0.5em, itemindent=2em}

\graphicspath{{images/}{Figures/}}  % add graphics path
\usepackage{xcolor} % required for color manipulation
\usepackage{cite} % merge consecutive numbers in \cite
\usepackage{soul}
\usepackage{comment}
\begin{document}
\title{Scattering-Mediated NLOS Ultraviolet Communication: \\ Channel Characterization, Modeling, and Data Transmission}

\author{Jian Shao,
        Tilahun Zerihun~Gutema, 
        and~Wasiu O.~Popoola% <-this % stops a space
        }%

%==========================================================================================
% make the title area
\maketitle

\begin{abstract}

Solar-blind ultraviolet (UV) optical wireless communication (OWC) enables non-line-of-sight (NLOS) transmission through atmospheric scattering. This work presents a comprehensive experimental and simulation study of a 100~m outdoor UVC link with comparative measurements under LOS-blocked and open-field conditions. The results show that LOS-blocked scattering-mediated links exhibit substantially higher path loss and narrower bandwidth than corresponding open-field measurements, highlighting the impact of residual LOS components in practical diffused beam UVC experiments. A photon-tracing Monte Carlo (MC) framework incorporating realistic transmitter/receiver characteristics and obstacle geometry is further developed to reproduce the measured channel characteristics with high accuracy. Despite the severe path loss and pulse broadening effects, reliable 20--200~kbps transmission is experimentally demonstrated over the 100~m LOS-blocked link using on-off keying (OOK) and pulse-position modulation (PPM), representing the longest-distance and highest-data-rate experimental demonstration of an explicitly LOS-blocked scattering-mediated UVC link reported to date.

\end{abstract}


% ========== Introduction Section ====================================
\section{Introduction}\label{section_introduction}

\IEEEPARstart {T}{he} field of optical wireless communications (OWC) has experienced rapidly growing research interest for next-generation communication technologies since the pioneering work of F.~R.~Gfeller and U.~Bapst in 1979 \cite{gfeller2005wireless}. This growth is driven by the availability of vast unregulated spectrum and the suitability of OWC in environments where radio-frequency (RF) signals are undesirable or propagate poorly. However, establishing stable and high-speed visible-light or infrared wireless links generally requires strict line-of-sight (LOS) conditions (\textit{i.e.} unobstructed propagation paths), mechanically stable transceivers, and benign atmospheric conditions. Otherwise, misalignment, vibration, turbulence, and blockage can severely deteriorate system performance or completely interrupt communication. To mitigate these challenges, specially deployed reflective structures and sophisticated beam-tracking systems are often required. These constraints are widely regarded as major limitations for the practical deployment of conventional OWC systems. Consequently, investigating the feasibility of outdoor scattering-mediated beyond-LOS OWC links, without dedicated reflective infrastructure, has become an important research topic.

Compared to visible-light and infrared communications, the development of ultraviolet (UV) communication has been relatively limited due to safety concerns. However, UV photons at 200--280~nm experience significantly stronger Rayleigh scattering from atmospheric molecules and suspended particles than visible or infrared light, making them more suitable for non-line-of-sight (NLOS) propagation. In addition, solar radiation within this wavelength range is strongly absorbed by the ozone layer in the stratosphere, resulting in negligible background noise at the ground level \cite{shaw2006extending}. Consequently, 200--280~nm ultraviolet optical wireless communication (UVC), also known as solar-blind UVC, enables photon-counting detection of ultra-weak signals even under direct sunlight. These unique physical characteristics make UVC highly promising for outdoor NLOS OWC applications \cite{shubin2020overview}.

In contrast to LOS or reflection-assisted NLOS links, scattering-mediated NLOS UVC links can even benefit from poor visibility weather conditions due to the increased density of scattering particulates \cite{shan2023non}. Furthermore, the IEEE 802.15.7 standard defines low-rate modes for OWC -- typically operating from 11.67 to 266.6~kb/s, for low-power, low-complexity, and wide-coverage applications such as intelligent transportation system (ITS) beaconing, positioning, and public information broadcasting. These unique characteristics suggest that scattering-mediated NLOS UVC systems have strong potential for communication in challenging environments, including the heavy-industry Internet of Things (IoT), covert battlefield communication, and disaster-relief operations \cite{guo2021ultraviolet}.

Currently, extensive analytical and numerical studies have been conducted on scattering channels, ranging from single-scattering analytical models to multi-scattering Monte Carlo (MC) simulations and analytical approximations \cite{luettgen1991non, drost2011uv, yuan2019monte, gong2016optical}. These studies have further investigated path loss and pulse broadening effects under various link geometries and atmospheric conditions \cite{zhang2024channel, elshimy2011non, shan2023non, ma2024effects}. Moreover, advances in UV optoelectronic devices, such as UV LEDs, solar-blind optical filters, and photomultiplier tubes (PMTs), have enabled a range of experimental demonstrations in the solar-blind UV spectrum \cite{shao2025enhanced, du2024implementation, shaw2006extending, li2023transmission}. Field tests have reported link path loss, pulse broadening, and data transmission performance under different atmospheric and geometric conditions \cite{chen2010experimental, zhao2024investigation, cao2024theoretical}.

However, the distinction between wide-beam LOS link and scattering-mediated NLOS link has not always been explicitly considered in existing experiments. Typical UVC setups often employ highly-diffused UV transmitters together with wide-angle and extremely-sensitive PMT receivers, while strict LOS blockage is not necessarily introduced \cite{he2024demonstration}. As a result, the received photons may contain contributions from both direct LOS and atmospheric scattered NLOS paths \cite{wu2024modeling}. Importantly, Drost \textit{et al.}~\cite{drost2013ultraviolet} numerically demonstrated that, for a Gaussian beam profile, the direct LOS component can dominate the received signal at low to medium transmitter/receiver (Tx/Rx) elevation angles, although its relative contribution decreases as the elevation angles increase. Such LOS-assisted propagation would be highly sensitive to physical blockage and can be readily suppressed once an obstacle occludes the direct LOS path, thereby reducing the effective obstacle-bypassing capability of the link. However, detailed calibration of Gaussian beam profiles and detector angular responses, as well as comparative investigations with and without explicit LOS blockage, remain relatively unexplored. Based on these considerations, this work presents a comprehensive study of scattering-mediated NLOS UVC through outdoor measurements and Monte Carlo simulations. The main contributions are summarised as follows:

\begin{itemize}[leftmargin=1em]

    \item \textbf{Channel characterisation:} A solar-blind, photon-level sensitive UVC testbed incorporating a physical LOS blocker is developed for rigorous scattering-mediated NLOS channel measurements. Comparative path-loss measurements with and without LOS blockage are conducted under multiple Tx/Rx elevation angles, providing insight into the influence of residual LOS propagation in UV links. In addition, a photon-accumulation-based frequency-response measurement method is developed to characterise scattering-mediated NLOS channels, enabling accurate bandwidth estimation from sparse photon-level signals.

    \item \textbf{Link simulation:} A multi-scattering Monte Carlo (MC) simulation framework is developed to reproduce the experimental link geometry and measurement conditions, incorporating realistic Tx/Rx characteristics, LOS propagation, and ground reflection. The simulation results show good agreement with the experimental measurements and demonstrate improved accuracy compared with conventional analytical models, providing a realistic framework for UVC channel analysis and system evaluation.

    \item \textbf{Data transmission:} Outdoor experiments demonstrate reliable 20--200~kbps data transmission using on-off keying non-return-to-zero (OOK-NRZ) and 4-pulse-position modulation (4-PPM) signalling over a 100~m scattering-mediated NLOS link under varying Tx/Rx elevation angles, representing a state-of-the-art performance among experimentally demonstrated obstacle-blocked UVC links.
    
\end{itemize}    

The remainder of this paper is organised as follows. Section~\ref{section_link_geometry} briefly discusses the differences between conventional NLOS scattering models and practical UV photon propagation behaviour, and reviews representative experimental link configurations and reported system performance in the literature. Section~\ref{section_experiment} presents the outdoor channel characterisation experiments, including measurement procedures, system path-loss results, and frequency-response characterisation. Section~\ref{section_simulation} describes the photon-tracing MC simulation framework and compares the results with both experimental measurements and a conventional analytical model. Section~\ref{section_data_transmission} demonstrates the data transmission performance of the UVC testbed. Finally, Section~\ref{section_conclusion} summarises the key findings and concludes the paper.

% ========== Theoretical Discussion ====================================

\section{Link Geometry and State-of-the-Art}\label{section_link_geometry}

\subsection{Photon Propagation Models}\label{subsection_uvppm}

\begin{figure}[!htbp]
    \centerline{\includegraphics[width=0.5\textwidth, keepaspectratio=true]{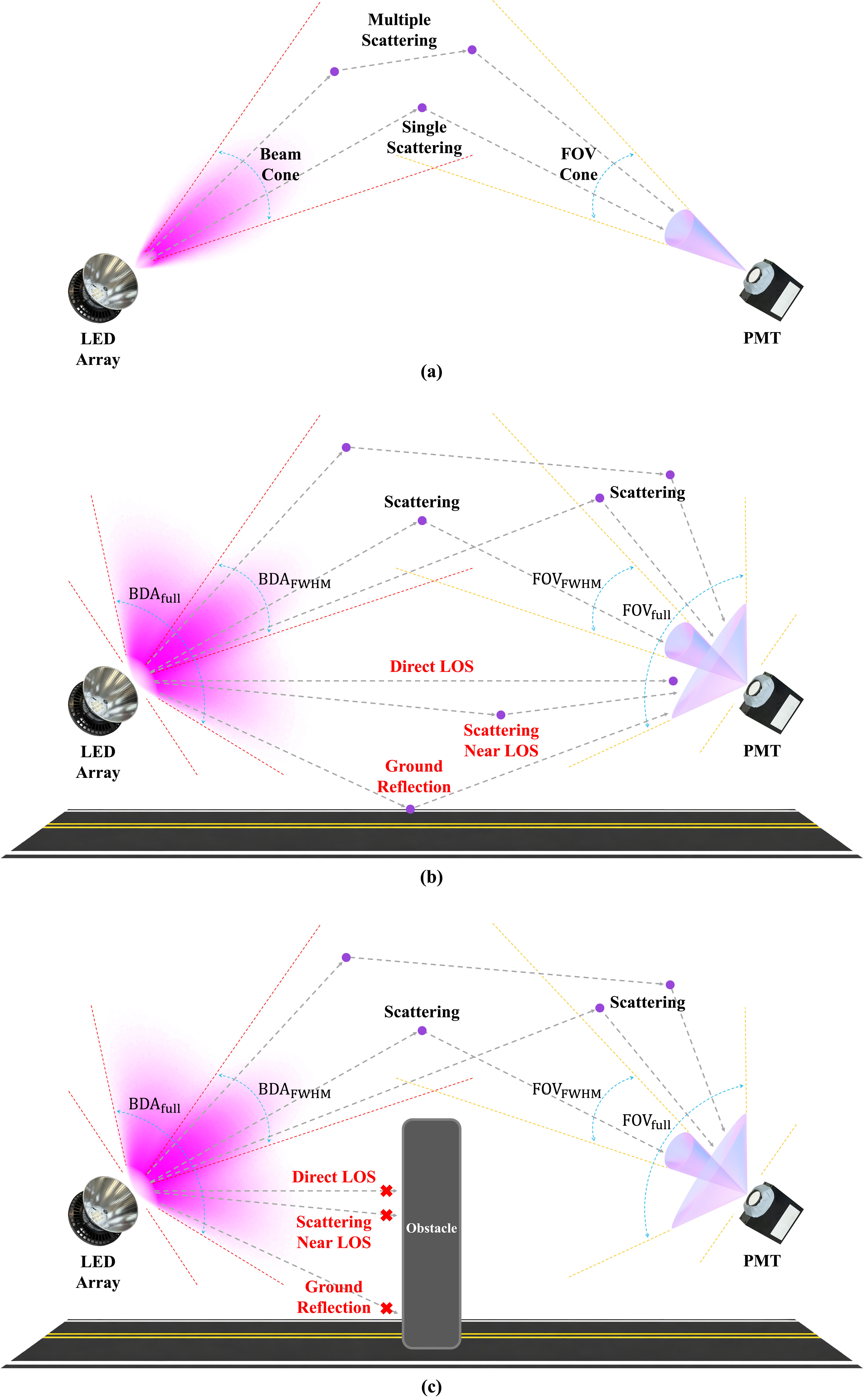}}
    \caption{Simplified theoretical model and practical experimental conditions: (a) conventional scattering-NLOS UVC model, (b) practical implementation conditions, and (c) the obstacle-bypassing scattering-mediated NLOS UVC channel.}
    \label{figure_sm}%
\end{figure}

Figure~\ref{figure_sm} illustrates the conventional scattering channel model shown in Fig.~\ref{figure_sm}(a), together with practical experimental link conditions without and with a physical obstacle in Fig.~\ref{figure_sm}(b) and Fig.~\ref{figure_sm}(c), respectively. 

Conventional theoretical models commonly approximate the Tx radiation pattern and Rx acceptance region as idealised BDA and FOV cones with hard boundaries, respectively \cite{song2020noise, wu2014experimental}. Under this assumption, direct LOS propagation is considered absent once the Tx/Rx pointing geometry falls outside the assumed maximum BDA/FOV boundaries, and the received signal is attributed solely to atmospheric scattering. This approximation is generally more appropriate for collimated narrow-beam sources such as lasers \cite{wu2025modeling}.

However, in practical UVC systems, LEDs and mercury lamps typically exhibit widely diffused Lambertian radiation patterns: $I(\theta)=I_0\cos^n(\theta)$, $0^\circ \le \theta \le 90^\circ$, where $I_0$ is the maximum radiant intensity, $n$ is the Lambertian emission order, and $\theta$ is the emission angle measured from the Tx pointing direction. Similarly, PMT detectors without dedicated front-end optical concentrators generally possess very broad angular acceptance characteristics. Since the complete beam pattern and detector acceptance region can extend over nearly the entire hemisphere ($\pm90^{\circ}$), manufacturer-reported angular specifications are commonly expressed using full-width-at-half-maximum (FWHM) values as practical indicators of the BDA and FOV. However, the FWHM merely denotes the angular position at which the optical intensity decreases to 50\% of its peak value rather than the maximum propagation boundary. Consequently, non-negligible optical power and detector sensitivity still exist beyond the nominal FWHM-defined limits. Combined with the high sensitivity of photon-counting PMT receivers, optical signals can therefore be transmitted and detected outside the nominal FWHM-defined BDA/FOV ranges. As illustrated in Fig.~\ref{figure_sm}(b), certain amount of LOS propagation may remain even when the Tx/Rx elevation angles exceed the nominal BDA/FOV values, such that the received signal can contain contributions from both direct-LOS and scattering-mediated propagation.

Therefore, while beam-cone and FOV-cone assumptions remain useful for simplified analysis and simulations, additional care is required when interpreting practical experiments employing diffused light sources and wide-angle high-sensitivity detectors. Without explicitly blocking the LOS path, it will be impossible to fully distinguish scattering-mediated photons from residual LOS components. In addition, obstacles can block not only the direct LOS path, but also nearby low-angle scattering paths and potential ground reflections. As illustrated in Fig.~\ref{figure_sm}(c), the introduction of a physical blocker provides a straightforward approach to suppress LOS and ground-reflection components under arbitrary Tx/Rx pointing conditions, enabling a controlled assessment of the feasibility of scattering-mediated NLOS UVC.

It is worth clarifying that physical blockage is not strictly required for a channel to become scattering-dominated. Even without obstacles, sufficiently large Tx/Rx elevation angles can reduce the LOS coupling so that atmospheric scattering becomes dominant \cite{drost2013ultraviolet}. Nevertheless, scattering-mediated NLOS UVC is primarily motivated by obstacle-bypassing communication scenarios where direct LOS paths are unavailable. In situations where LOS propagation remains possible through appropriate Tx/Rx alignment, the resulting link would generally offer substantially lower path loss, larger channel bandwidth, and higher achievable data rates. Therefore, explicit blockage of the LOS path provides a more realistic and application-oriented framework for evaluating obstacle-bypassing communication scenarios.

\subsection{State-of-the-Art}\label{subsection_sota}

\begin{table*}[!t]
\caption{Reported channel-characterization experiments for scattering-mediated NLOS UVC links}
\label{table_lrcc}
\centering
\tiny
\setlength{\tabcolsep}{2.5pt}
\renewcommand{\arraystretch}{1.15}

\resizebox{\textwidth}{!}{%

\begin{tabular}{
>{\centering\arraybackslash}p{1.3cm}
>{\centering\arraybackslash}p{2.0cm}
>{\centering\arraybackslash}p{2.0cm}
>{\centering\arraybackslash}p{2.0cm}
>{\centering\arraybackslash}p{2.2cm}
>{\centering\arraybackslash}p{1.4cm}
>{\centering\arraybackslash}p{1.7cm}
>{\centering\arraybackslash}p{2.6cm}
>{\centering\arraybackslash}p{0.9cm}
}

\toprule

\multicolumn{3}{c}{\textbf{Link Geometry}} &
\multicolumn{3}{c}{\textbf{Hardware Configuration}} &
\multirow{2}{*}{\textbf{Environmental Condition}} &
\multirow{2}{*}{\textbf{Channel Characterization}} &
\multirow{2}{*}{\textbf{Reference}} \\

\cmidrule(lr){1-3}
\cmidrule(lr){4-6}

\textbf{Distance} &
\textbf{Blocking Condition} &
\textbf{Tx/Rx Elevation Angle} &
\textbf{FWHM BDA/FOV} &
\textbf{Light Source} &
\textbf{Photon Sensor} &
&
&
\\

\midrule

\multicolumn{9}{c}{\textit{Group 1: Scattering-NLOS UVC links without explicit LOS blockage}} \\

\midrule

10/20 m &
No LOS blockage &
(10--80\textdegree, 0--80\textdegree) &
$\pm$5\textdegree/$\pm$15\textdegree &
265 nm LED (20 mW) &
PMT + filter &
Clear atmosphere &
Path loss vs. elevation angle &
\cite{raptis2016power} \\

2--50 m &
No LOS blockage &
(0--30\textdegree, 0--50\textdegree) &
$\pm$3\textdegree/$\pm$15\textdegree &
265 nm LED (0.36 W) &
PMT only &
Nighttime &
Path loss vs. elevation angle and distance &
\cite{luo2012performance} \\

20--70 m &
No LOS blockage &
(0--45\textdegree, 0--45\textdegree) &
$\pm$5\textdegree/unspecified &
250/280 nm LED (60/84 mW) &
PMT + filter &
Unspecified &
Path loss vs. elevation/azimuth angle &
\cite{xu2007experimental} \\

% 10--100 m &
% No LOS blockage &
% (10--90\textdegree, 0--90\textdegree) &
% $\pm$0.086\textdegree/$\pm$15\textdegree &
% \makecell{266 nm laser + lens \\ ($\sim$1 MW pulses)} &
% PMT + filter + lens &
% Unspecified &
% Impulse response vs. elevation angle, distance and BDA/FOV &
% wd \\

25 m / 1--100 m &
Open field &
(0--90\textdegree, 0--90\textdegree) &
$\pm$5\textdegree/$\pm$15\textdegree &
250 nm LED (2.1 mW) &
PMT + filter &
Sunny day &
Path loss vs. elevation angle and distance &
\cite{chen2008experimental} \\

100/400 m &
Along a road &
(8--56\textdegree, 8\textdegree) &
$\pm$10\textdegree/$\pm$10\textdegree &
265 nm LED + lens (0.41 W) &
PMT + filter &
Afternoon &
Path loss vs. Tx elevation angle &
\cite{du2022characteristic} \\

% 500 m--4 km &
% Flat terrain &
% (10--90\textdegree, 10--70\textdegree) &
% $\pm$0.086\textdegree/$\pm$15\textdegree &
% \makecell{266 nm laser + lens \\ ($\sim$1 MW pulses)} &
% PMT + filter + lens &
% At night &
% Path loss vs. elevation angle and distance / impulse response &
% ds \\

\midrule

\multicolumn{9}{c}{\textit{Group 2: LOS-blocked or high-elevation-angle scattering-mediated NLOS UVC links}} \\

\midrule

10--30 m &
No LOS blockage &
(60--90\textdegree, 60--90\textdegree) &
$\pm$30\textdegree/$\pm$15\textdegree &
250/280 nm LED (50 mW) &
PMT + filter &
Nighttime &
Step response vs. elevation angle and distance &
\cite{cao2024theoretical} \\

40--400 m &
Building obstacle &
(60--110\textdegree, 10--130\textdegree) &
$\pm$10\textdegree/$\pm$26\textdegree &
254 nm mercury lamp + reflector (1 W) &
PMT + filter &
Afternoon &
Path loss vs. elevation angle and distance &
\cite{zhang2011study} \\

100 m &
3×5 m curtain &
(0--90\textdegree, 0--90\textdegree) &
$\pm$9\textdegree/$\pm$23.75\textdegree &
\makecell{265 nm LED + reflector \\ (0.9 W)} &
PMT + filter &
Nighttime (AQI $\approx$ 15) &
Path loss / frequency response vs. elevation angle &
\textbf{This Work} \\

\bottomrule

\end{tabular}

}
\end{table*}

\begin{table*}[!t]
\caption{Reported data-transmission experiments for scattering-mediated NLOS UVC links}
\label{table_lrsp}
\centering
\tiny
\setlength{\tabcolsep}{2.5pt}
\renewcommand{\arraystretch}{1.15}

\resizebox{\textwidth}{!}{%

\begin{tabular}{
>{\centering\arraybackslash}p{1.2cm}
>{\centering\arraybackslash}p{2.4cm}
>{\centering\arraybackslash}p{2.0cm}
>{\centering\arraybackslash}p{2.0cm}
>{\centering\arraybackslash}p{2.2cm}
>{\centering\arraybackslash}p{1.5cm}
>{\centering\arraybackslash}p{2.0cm}
>{\centering\arraybackslash}p{3.0cm}
>{\centering\arraybackslash}p{0.9cm}
}

\toprule

\multicolumn{3}{c}{\textbf{Link Geometry}} &
\multicolumn{3}{c}{\textbf{Hardware Configuration}} &
\multirow{2}{*}{\textbf{Environmental Condition}} &
\multirow{2}{*}{\textbf{System Performance}} &
\multirow{2}{*}{\textbf{Reference}} \\

\cmidrule(lr){1-3}
\cmidrule(lr){4-6}

\textbf{Distance} &
\textbf{Blocking Condition} &
\textbf{Tx/Rx Elevation Angle} &
\textbf{FWHM BDA/FOV} &
\textbf{Light Source} &
\textbf{Photon Sensor} &
&
&
\\

\midrule

\multicolumn{9}{c}{\textit{Group 1: Low-elevation-angle scattering-NLOS UVC links without explicit LOS blockage}} \\

\midrule

30 m &
No LOS blockage &
(10\textdegree, 10\textdegree) &
$\pm$3.5\textdegree/$\pm$15\textdegree &
265 nm LED (15.5 mW) &
PMT + filter &
Unspecified &
\makecell{500 kbps OOK-NRZ \\ (BER $\approx 3.5\times10^{-3}$)} &
\cite{wu2014experimental} \\

65 m &
\makecell{Sparse leafless trees \\ (3--4 m tall)} &
(20\textdegree, 30\textdegree) &
Unspecified &
254 nm mercury lamps &
PMT + filter &
Unspecified &
\makecell{5 kbps OOK-NRZ \\ (BER $=1.52\times10^{-3}$)} &
\cite{zhao2024investigation} \\

75.1 m &
No LOS blockage &
(30--50\textdegree, 30--50\textdegree) &
$\pm$8.5\textdegree/$\pm$15\textdegree &
\makecell{250 nm LED + reflector \\ (0.5 W)} &
PMT + filter &
Unspecified &
50 kbps OOK-NRZ &
\cite{li2024ultraviolet} \\

110 m &
No LOS blockage &
(30\textdegree, 10\textdegree) &
$\pm$60\textdegree/$\pm$15\textdegree &
266 nm LED &
PMT + filter &
Cloudy &
800 kbps OOK-NRZ &
\cite{pan2024beacon} \\

200 m &
No LOS blockage &
No misalignment &
$\pm$20\textdegree/$\pm$15\textdegree &
272 nm LED + lens (40 mW) &
PMT + filter &
\makecell{Clear weather \\ (76\% humidity)} &
\makecell{300 kbps OOK-NRZ audio} &
\cite{deng2021realization} \\

450/930 m &
Open grassland &
\makecell{(0\textdegree, 24\textdegree)/(8\textdegree, 24\textdegree)} &
$\pm$7.5\textdegree/$\pm$22.5\textdegree &
268 nm LED (480 mW) &
PMT + filter &
\makecell{Few clouds} &
\makecell{142.8/71.4 kbps 4-PPM} &
\cite{he2024demonstration} \\

% 1 km &
% \makecell{Placed on buildings} &
% $\sim$1\textdegree~misalignment &
% $\pm$0.023\textdegree/$\pm$22.5\textdegree &
% 266 nm laser (120 mW) &
% PMT + filter &
% Cloudy &
% 1.113 Mbps OOK-NRZ &
% wang \\

\midrule

\multicolumn{9}{c}{\textit{Group 2: LOS-blocked or high-elevation-angle scattering-mediated NLOS UVC links}} \\

\midrule

11 m &
No LOS blockage &
(90\textdegree, 90\textdegree) &
Unspecified &
274 nm LED (40 mW) &
PMT + filter &
Full sunlight &
2.4 kbps 4-PPM audio &
\cite{shaw2006extending} \\

50 m &
Car obstacle &
(0--65\textdegree, 0--65\textdegree) &
$\pm$10\textdegree/$\pm$10\textdegree &
266 nm LED + lenses (1.3 W) &
PMT + filter &
Urban air (AQI $\approx$ 70) &
\makecell{10 kbps OOK-NRZ \\ (BER $\le 2.2\times10^{-3}$)} &
\cite{du2024implementation} \\

100 m &
3×5 m curtain &
(0--90\textdegree, 0--90\textdegree) &
$\pm$9\textdegree/$\pm$10\textdegree &
\makecell{265 nm LED + reflector \\ (1.8 W)} &
PMT + filter + concentrator &
Afternoon (AQI $\approx$ 15) &
\makecell{20--200 kbps OOK-NRZ/4-PPM} &
\textbf{This Work} \\

\bottomrule

\end{tabular}

}
\end{table*}

Representative atmospheric-scattering NLOS UVC channel-characterisation and data-transmission experiments reported in the literature are summarised in Table~\ref{table_lrcc} and Table~\ref{table_lrsp}, respectively. The listed parameters include link geometry, hardware configuration, atmospheric conditions, and experimental performance. For comparison, the studies are divided into two groups according to whether explicit LOS blockage was introduced or whether sufficiently large Tx/Rx elevation angles were employed.  

Most reported experiments employed naked UV LEDs or mercury lamps (with or without reflectors) together with naked PMT detectors, resulting in widely diffused Lambertian beams and wide-angle photon-level reception ability. In Table~\ref{table_lrcc}, the majority of Group~1 channel characterisation experiments started from very low Tx/Rx elevation angles, including no misalignment conditions, and gradually increased the pointing angles. Without explicit LOS blockage, the measured links therefore contain varying contributions from both diffused LOS and atmospheric-scattering components. Similarly, in Table~\ref{table_lrsp}, several Group~1 data transmissions were conducted with Tx/Rx elevation angles within or only slightly beyond the nominal FWHM BDA/FOV ranges. Under open-field conditions without physical blockage, residual LOS coupling may still contribute to the measured system performance. 

As a result, experimental investigations focusing specifically on obstacle-bypassing scattering-mediated NLOS UVC links remain relatively limited. In particular, comprehensive comparative measurements with and without LOS blockage are still scarce. Motivated by this, the present work performs rigorous path-loss and frequency-response characterisation of scattering-mediated NLOS UVC channels and demonstrates high-performance data transmission under explicitly LOS-blocked conditions over a 100~m outdoor link.

% ========== Experiment ====================================

\section{Outdoor Channel Measurement and Characterisation}\label{section_experiment}

\subsection{Experimental Setups and Configurations}\label{subsection_esuac}

\begin{figure}[!htbp]
    \centerline{\includegraphics[width=0.5\textwidth, keepaspectratio=true]{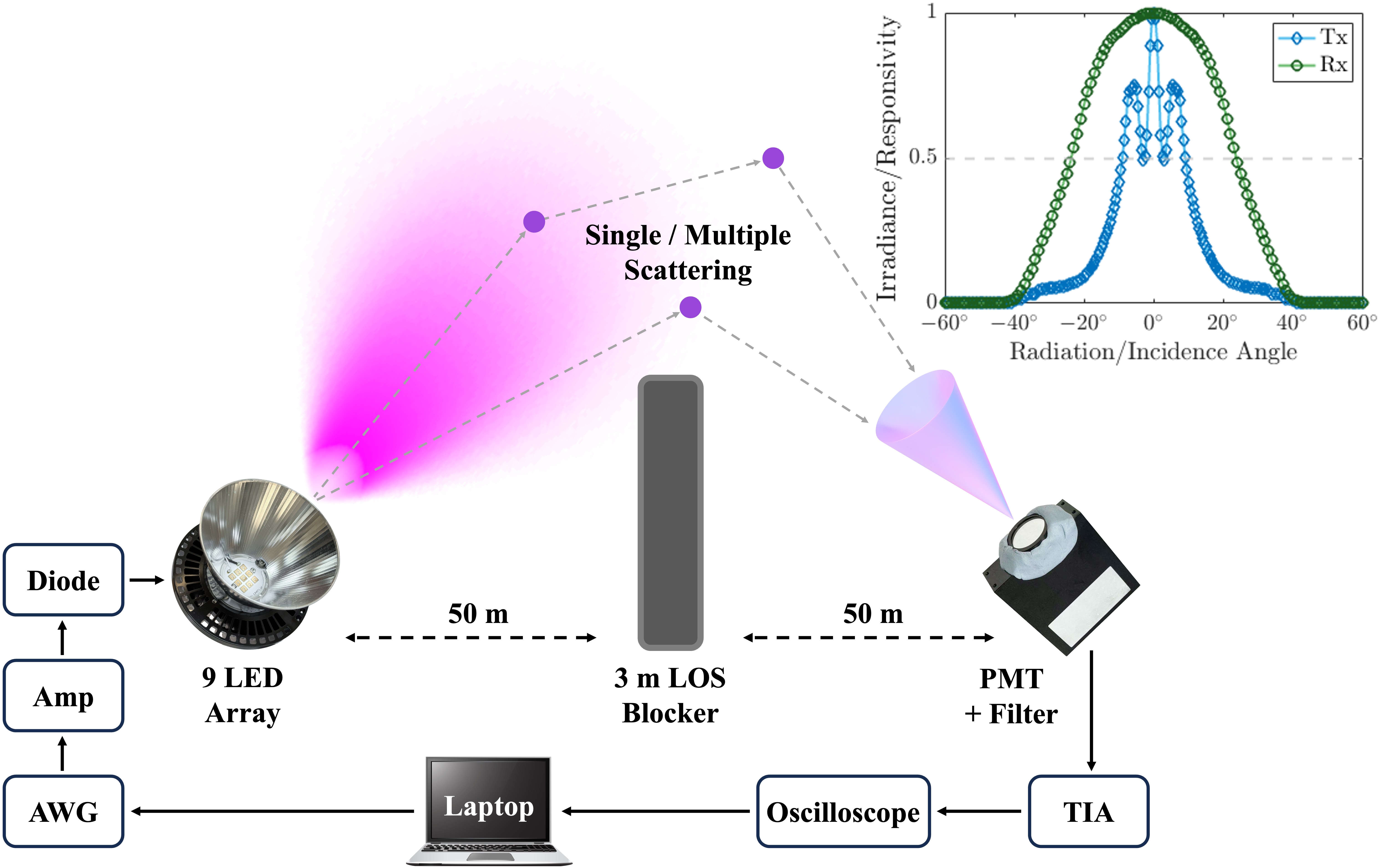}}
    \caption{Photon-counting scattering NLOS UVC system block diagram. \\
    (Inset: Measured beam pattern and receiver responsivity curve.)}
    \label{figure_sbd}%
\end{figure}

\begin{figure}[!htbp]
    \centerline{\includegraphics[width=0.45\textwidth, keepaspectratio=true]{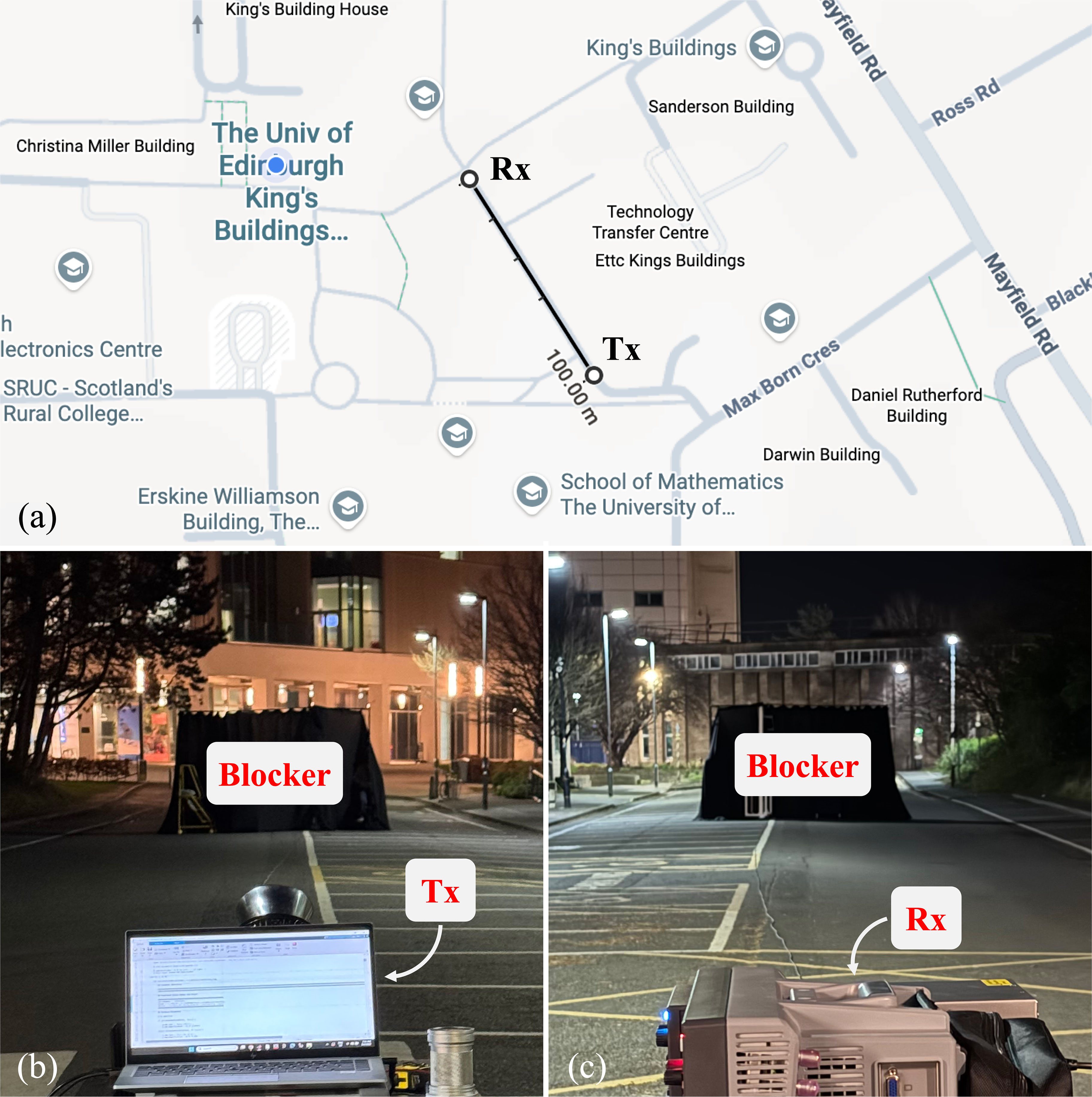}}
    \caption{Outdoor experimental environment: (a) Tx and Rx location on the map. (b) View from Tx side. (c) View from Rx side.}
    \label{figure_sps}%
\end{figure}

Figure~\ref{figure_sbd} presents the block diagram of the proposed solar-blind photon-level UVC testbed for outdoor measurements, while photographs of the experimental environment are shown in Fig.~\ref{figure_sps}. The system performs pulsed-signal transmission and photon-counting detection using a PMT-based receiver.

For path-loss measurements, the transmitter (Tx) employs an arbitrary waveform generator (AWG) to generate bipolar square-waves. The electrical signal is amplified by an OPA541 power amplifier before driving the UV LEDs. A fast-switching Schottky diode (SR240) clips the negative voltage to prevent reverse breakdown of the LEDs. The optical source consists of a $3\times3$ array of 265~nm UV LEDs (SRAM SU CZHEF1.VC), providing up to 0.9~W peak optical power. A spotlight reflector (F13662\_ANGELA-S-B) is attached to improve beam directivity, with an estimated optical loss of approximately 15\% due to the aluminium reflective surface.  

At the receiver (Rx), the ultra-weak UV signals are detected using a PMT (Hamamatsu H9305-09) operating in high-gain photon-counting mode. An OD4 optical bandpass filter (Newport 10BPF10-26) suppresses out-of-band ambient light while providing about 12\% transmittance for 265~nm photons at normal incidence angle. The measured Tx beam pattern and Rx angular responsivity are shown in the inset of Fig.~\ref{figure_sbd}. The FWHM BDA/FOV values are $\pm 9^{\circ}$ and $\pm 23.75^{\circ}$, respectively. However, the corresponding angular ranges at the 1\% intensity/responsivity level further extend to $\pm 40^{\circ}$ and $\pm 41^{\circ}$, illustrating the highly diffused transmission and wide-angle reception characteristics of the hardware. Approximately 20\% of incident photons generate detectable photocurrent spikes in the PMT. These weak electrical pulses are further amplified by a transimpedance amplifier (TIA) (Hamamatsu C11184) and captured by the oscilloscope for offline processing. A laptop controls signal transmission and stores the measured data.  

As shown in Fig.~\ref{figure_sps} (a), the Tx and Rx benches are separated by 100~m in an outdoor car-park on campus, with the separation distance measured and verified using a laser rangefinder. A thick black curtain with dimensions of $3\times5$~m is positioned midway between the Tx and Rx to fully block the direct LOS path and significantly suppress ground reflections. Both the LED array and PMT are mounted 1~m above the ground, while the blocker height exceeds the Tx/Rx LOS path by at least 2~m. The key experimental parameters are summarised in Table~\ref{table_kpoes}.

\begin{table}[!htbp]
\centering
\caption{\label{table_kpoes} Key Parameters of Experimental Setup.\\}
\begin{tabular}{|c|c|}
\hline
\textbf{Parameter} & \textbf{Value} \\ 
\hline
\hline
Tx-Rx Distance $d_{\textup{Tx-Rx}}$ & $100~\textup{m}$ \\
\hline
Tx/Rx-Obstacle Distance $d_{\textup{Tx-obst}}$/$d_{\textup{Rx-obst}}$ & $50~\textup{m}$ \\
\hline
Bench Height $h_{\textup{Tx}}$/$h_{\textup{Rx}}$ & $1~\textup{m}$ \\
\hline
Obstacle Size $h_{\textup{obst}}$/$w_{\textup{obst}}$ & $3~\textup{m}$ tall $\times$ $5~\textup{m}$ wide\\
\hline
Tx/Rx Elevation Angle $\theta_{\textup{Tx}}$/$\theta_{\textup{Rx}}$ & $0^{\circ}, 15^{\circ}, 30^{\circ}, ..., 90^{\circ}$ \\
\hline
LED Wavelength $\lambda$ & $265~\textup{nm}$ \\
\hline
LED Array Peak Power $P_{\textup{peak}}$ & $0.9~\textup{W}$ \\
\hline
Estimated Reflector Efficiency $\eta_{\textup{reflector}}$ & $85\%$ \\
\hline
Filter Transmittance $T_{\textup{filter}}$ (at $0^{\circ}$~AOA) & $12\%$ \\
\hline
PMT Sensing Area $A_{\textup{PMT}}$ & $0.481~\textup{cm}^{2}$ ($0.37 \times 1.3~\textup{cm}$) \\
\hline
PMT Detection Efficiency $\eta_{\textup{PMT}}$ & $20\%$ \\
\hline
$\textup{BDA}_{\textup{FWHM}}$ / $\textup{FOV}_{\textup{FWHM}}$ & $\pm 9^{\circ}$ / $\pm 23.75^{\circ}$ \\
\hline
$\textup{BDA}_{1\%}$ / $\textup{FOV}_{1\%}$ & $\pm 40^{\circ}$ / $\pm 41^{\circ}$ \\
\hline
\end{tabular}
\end{table}

\subsection{Path Loss Measurement}\label{subsection_plm}

To reduce the influence of ambient light on the measurement resolution and enable accurate path-loss estimation at extremely weak received power levels, the outdoor experiment was conducted during nighttime after sunset on 04/03/2026. The background photon counting rate (PCR), including ambient-light leakage and PMT dark counts, was measured as 574~counts/s. During the measurement, the weather was clear with a temperature of approximately $8^{\circ}$C, a relative humidity of 42\%, and an excellent air quality index (AQI) of around 15.  

Without loss of generality, a co-planar Tx/Rx geometry was considered, where no azimuthal misalignment was introduced between the Tx and Rx. The Tx/Rx elevation angles, denoted by $\theta_{\mathrm{Tx}}$ and $\theta_{\mathrm{Rx}}$, were varied from $0^{\circ}$ to $90^{\circ}$ in steps of $15^{\circ}$, resulting in 49 elevation-angle combinations. For compact notation, Tx$_{n^{\circ}}$ and Rx$_{n^{\circ}}$ denote $\theta_{\mathrm{Tx}}=n^{\circ}$ and $\theta_{\mathrm{Rx}}=n^{\circ}$, respectively. In some figures, $n^{\circ}$ is further used as a shorthand for the symmetric case of $\theta_{\mathrm{Tx}}=\theta_{\mathrm{Rx}}=n^{\circ}$.  

For each measurement, the Tx emitted a 1~kHz square-wave signal, periodically switching the LED array between the on and off states. The oscilloscope sampled the received waveform at 1~GSa/s over a duration of 100~ms, and each measurement point was repeated five times. Since the PMT operated in photon-counting mode, each detected photon generates a distinguishable photocurrent spike. The photon-counting rate (PCR) $R$ is therefore calculated as
\begin{equation}
R = \frac{N_{\mathrm{p}}}{T_{\mathrm{d}}},
\label{eq:pcr}
\end{equation}
where $N_{\mathrm{p}}$ is the number of detected photons within the detection duration $T_{\mathrm{d}}$. 

The measured PCR during the LED on state contains both received UV signal photons and background counts caused by ambient light leakage and PMT dark counts. However, the PCR measured during the LED off state solely corresponds to the background counts. Therefore, the pure received UV signal PCR is obtained as
\begin{equation}
R_{\mathrm{sig}} = R_{\mathrm{on}} - R_{\mathrm{off}},
\label{eq:signal_pcr}
\end{equation}
where $R_{\mathrm{on}}$ and $R_{\mathrm{off}}$ denote the measured PCRs during the LED on and off states, respectively.  

The energy of a single UV photon is given by
\begin{equation}
E_{\mathrm{p}} = \frac{h c}{\lambda},
\label{eq:photon_energy}
\end{equation}
where $h$ is Planck's constant, $c$ is the speed of light in vacuum, and $\lambda=265~\mathrm{nm}$ is the wavelength. 

Then, the received UV power can be estimated as
\begin{equation}
P_{\mathrm{r}}
= R_{\mathrm{sig}} E_{\mathrm{p}}
= \frac{h c \left(R_{\mathrm{on}}-R_{\mathrm{off}}\right)}{\lambda},
\label{eq:rx_optical_power}
\end{equation}
where $P_{\mathrm{r}}$ denotes the received signal optical power.  

Finally, the system total path loss in dB is calculated from the ratio between the transmitted and received optical powers:
\begin{equation}
L_{\mathrm{total}}
= 10\log_{10}\left(\frac{P_{\mathrm{emit}}}{P_{\mathrm{r}}}\right)
= 10\log_{10}\left(
\frac{P_{\mathrm{peak}}\lambda}
{h c \left(R_{\mathrm{on}}-R_{\mathrm{off}}\right)}
\right).
\label{eq:total_path_loss}
\end{equation}

\begin{figure}[!htbp]
    \centerline{\includegraphics[width=0.5\textwidth, keepaspectratio=true]{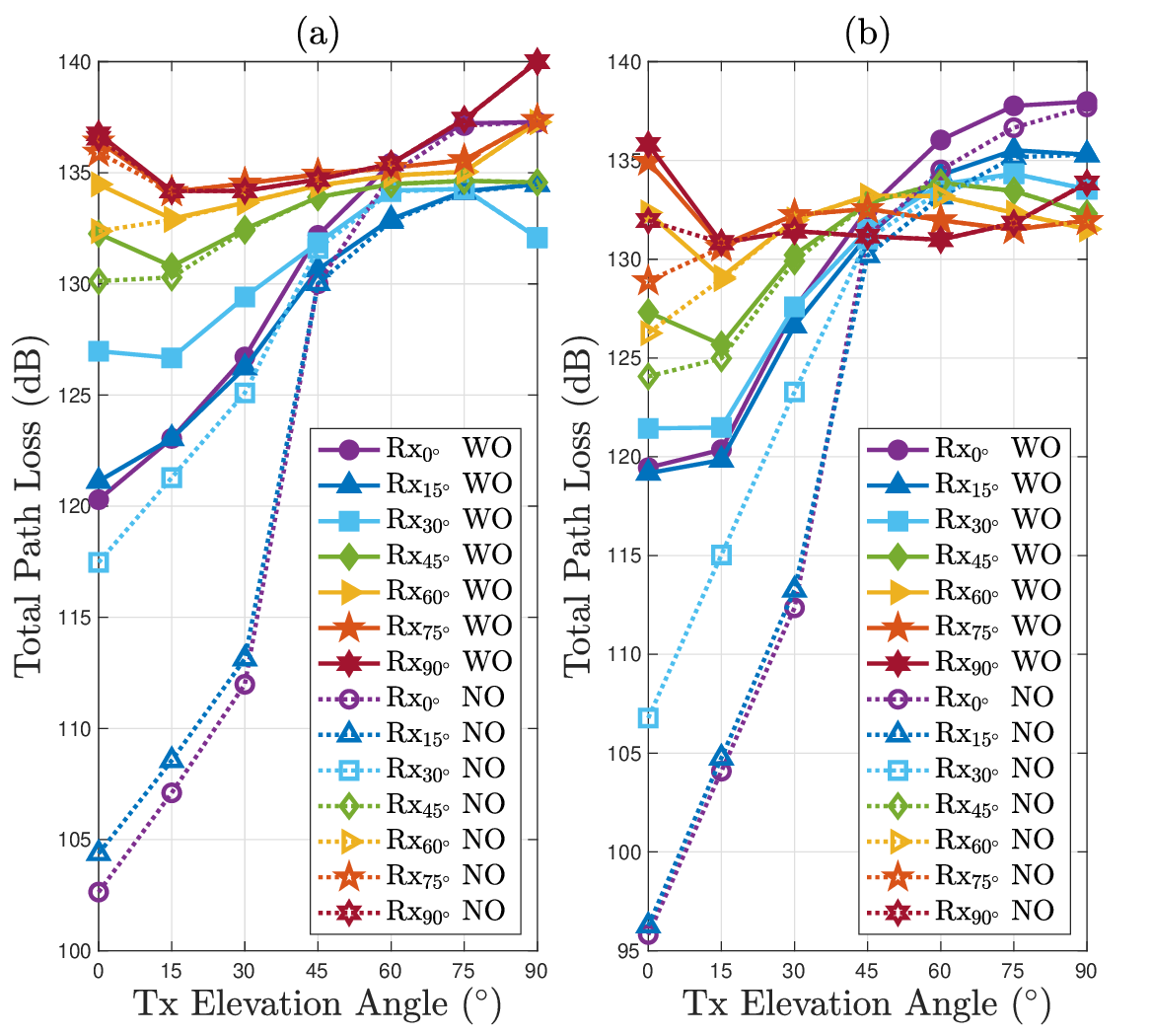}}
    \caption{Total path loss: (a) Experimental measurements. (b) Simulation results. (WO: with obstacle / NO: no obstacle)}
    \label{figure_tples}%
\end{figure}

After completing the measurements under strict LOS blockage conditions, the obstacle was removed and a reference experiment without LOS blockage was conducted. The measured path loss for both cases are presented in Fig.~\ref{figure_tples}(a). The solid lines correspond to the with-obstacle (WO) measurements, while the dotted lines represent the no-obstacle (NO) measurements. Each line shows the total path loss as the $\theta_{\mathrm{Tx}}$ increases from $0^{\circ}$ to $90^{\circ}$ with a fixed $\theta_{\mathrm{Rx}}$.

Comparing the WO and NO datasets, clear differences can be observed after removing the LOS blocker. For example, under the Tx$_{15^{\circ}}$/Rx$_{30^{\circ}}$ geometry, the $\theta_{\mathrm{Tx}}$/$\theta_{\mathrm{Rx}}$ already exceed the $\textup{BDA}_{\mathrm{FWHM}}$/$\textup{FOV}_{\mathrm{FWHM}}$ limits of $\pm9^{\circ}$/$\pm23.75^{\circ}$. According to the conventional criterion, this configuration would typically already be regarded as a pure atmospheric-scattering NLOS link. However, after introducing the LOS blocker, the measured path loss increases by about 5.4~dB, indicating that non-negligible LOS-related signal components are still present beyond the nominal FWHM boundaries. Similar behaviour was observed across multiple elevation-angle geometries.

Furthermore, the overall trend of the NO measurements is similar to other studies such as Fig.~5 in~\cite{chen2008experimental} and Fig.~3 in~\cite{raptis2016power}, where potential diffused-LOS components are also likely present. In the NO case, lower $\theta_{\mathrm{Tx}}$/$\theta_{\mathrm{Rx}}$ generally correspond to stronger direct LOS contributions and lower path losses. After introducing the LOS blocker to explicitly block the direct LOS path, the path loss at lower elevation angles increases significantly, revealing the much weaker underlying atmospheric-scattering channel.

Two characteristic features can also be observed in the WO measurements. First, when the $\theta_{\mathrm{Tx}}$ increases from $0^{\circ}$ to $15^{\circ}$, the path loss slightly decreases in some cases. This is likely because more photons can bypass the obstacle instead of being blocked or absorbed by the obstacle and ground, allowing more photons to participate in atmospheric scattering. Second, as the $\theta_{\mathrm{Tx}}$ further increases, the longer propagation distance and weaker large-angle scattering both increase the path loss. However, the overlap volume between the beam and the FOV also increases significantly at higher elevations, enhancing the probability that scattered photons are seen by the Rx. Consequently, the path loss increase gradually slows down and may even slightly reduce at medium-to-high elevation angles. The corresponding Monte Carlo (MC) simulation results are presented in Fig.~\ref{figure_tples}(b), where similar trends can also be observed. More detailed discussions will be provided in Section~\ref{section_simulation}.

\subsection{Frequency Response Measurement}\label{subsection_frm}

\begin{figure}[!htbp]
    \centerline{\includegraphics[width=0.515\textwidth, keepaspectratio=true]{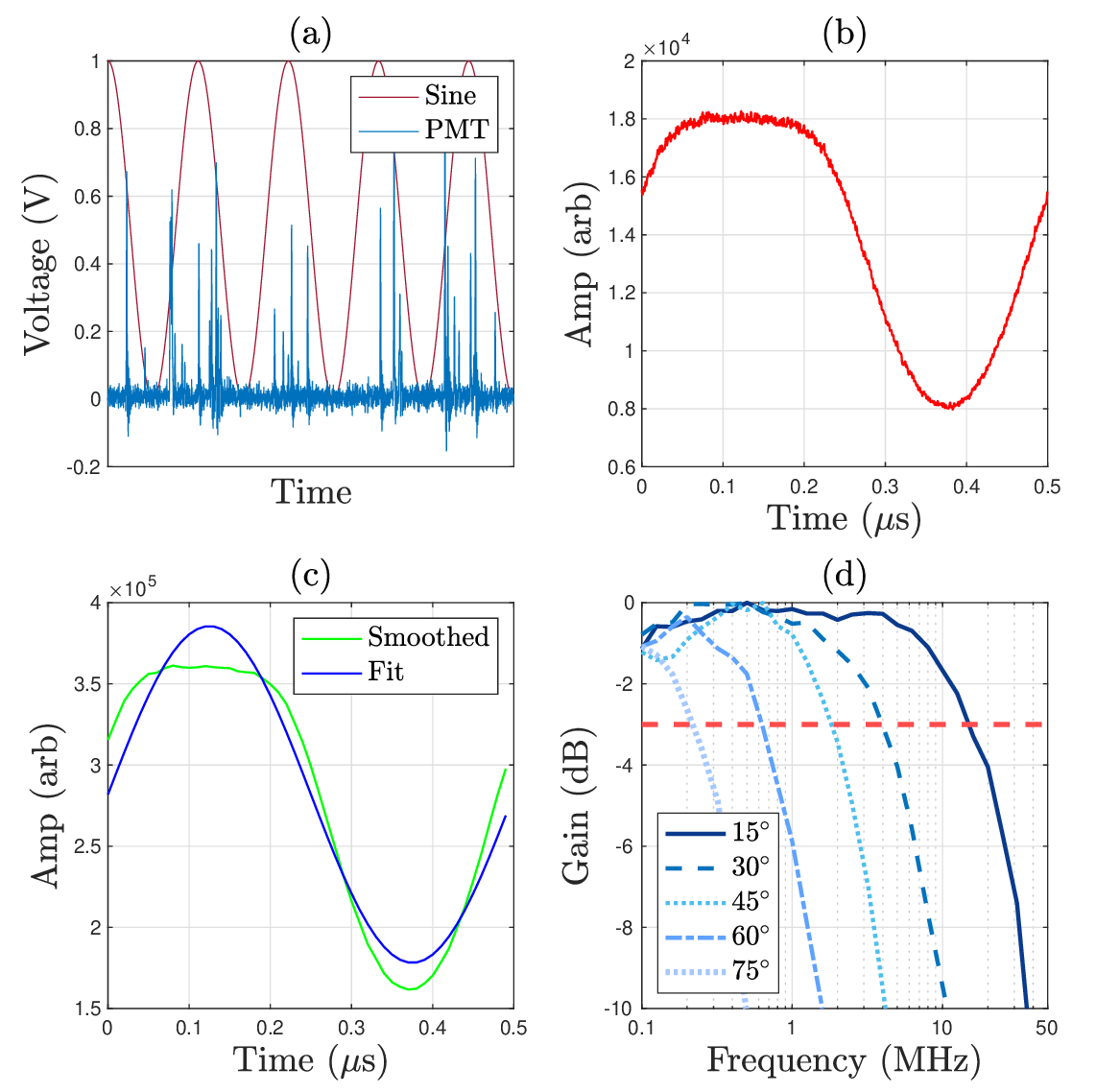}}
    \caption{Frequency response measurement using a sine-wave sweep method: (a) Sine wave reference and PMT output. (b) Superimposed sine wave. (c) Smoothed and fitted sine wave. (d) Frequency responses for varying Tx-Rx elevation angles.}
    \label{figure_sfr}%
\end{figure}

Since atmospheric-scattering channels comprise numerous propagation paths with different travelling distances, severe pulse broadening is expected, resulting in a reduced channel bandwidth and consequently limiting the achievable data rate. Therefore, accurate channel-bandwidth characterisation is an important performance metric for scattering channels. In common practice, channel bandwidth is obtained using sine-wave frequency sweeping, where the channel frequency response is determined from the gain of received sinusoidal signals at different frequencies \cite{broch1975measurement}. However, the extremely high path loss of scattering-mediated NLOS UVC links, often exceeding 120~dB, makes direct continuous-time sinusoidal waveform measurements with a PMT receiver impractical. When a sinusoidal signal is transmitted, the received waveform consists only of sparse photocurrent spikes generated by individual detected photons. As illustrated in Fig.~\ref{figure_sfr}(a), the spikes become denser near the peaks of the sine wave and sparser near the valleys. Consequently, the received sinusoidal waveform peak-to-peak voltage ($V_{\mathrm{pp}}$) cannot be directly measured.

To address this issue, repeated received waveforms were periodically superimposed. As shown in Fig.~\ref{figure_sfr}(b), after sufficient accumulation, the discrete photocurrent spikes gradually form a continuous sinusoidal waveform. Slight distortion can be observed near the sine-wave peak due to LED nonlinearity under relatively large driving $V_{\mathrm{pp}}$. Although smaller input $V_{\mathrm{pp}}$ could reduce distortion, it would also lead to insufficient SNR and inaccurate gain estimation. Therefore, the LED driving $V_{\mathrm{pp}}$ was chosen as a compromise between waveform linearity and measurement resolution. In addition, small unavoidable timing mismatches between the AWG and oscilloscope at certain frequencies produced minor sawtooth-like fluctuations in the accumulated waveform. To suppress this influence, every $n$ adjacent samples of the accumulated waveform in Fig.~\ref{figure_sfr}(b) were further summed, resulting in the smoothed signal shown in Fig.~\ref{figure_sfr}(c). The processed waveform was then fitted using a sinusoidal wave with known frequency to estimate the received $V_{\mathrm{pp}}$.  

The measured $V_{\mathrm{pp}}$ values at different frequencies were normalised according to the accumulation cycles and smoothing parameters, producing the frequency-response curves shown in Fig.~\ref{figure_sfr}(d). During the experiment, the OPA541 amplifier and Schottky diode were replaced with a Mini-Circuits amplifier (ZHL-6A-S+) and a Tektronix PSPL5575A bias-tee to enable sinusoidal modulation of the LED array. The Tx/Rx elevation angles were jointly varied from $15^{\circ}$ to $75^{\circ}$ under the condition $\theta_{\mathrm{Tx}}=\theta_{\mathrm{Rx}}$. The extracted bandwidth values are summarised in Fig.~\ref{figure_cirbw}(b) together with the simulation results. The measured channel bandwidth gradually decreases as $\theta_{\mathrm{Tx}}$ increases, indicating progressively stronger scattering-induced multipath broadening at higher elevation angles.

% ========== Simulation ====================================
\section{Photon Tracing-based Monte Carlo Simulation}\label{section_simulation}

\subsection{Link Geometry and Simulation Procedure}\label{subsection_lgasp}

\begin{figure}[!htbp]
    \centerline{\includegraphics[width=0.505\textwidth, keepaspectratio=true]{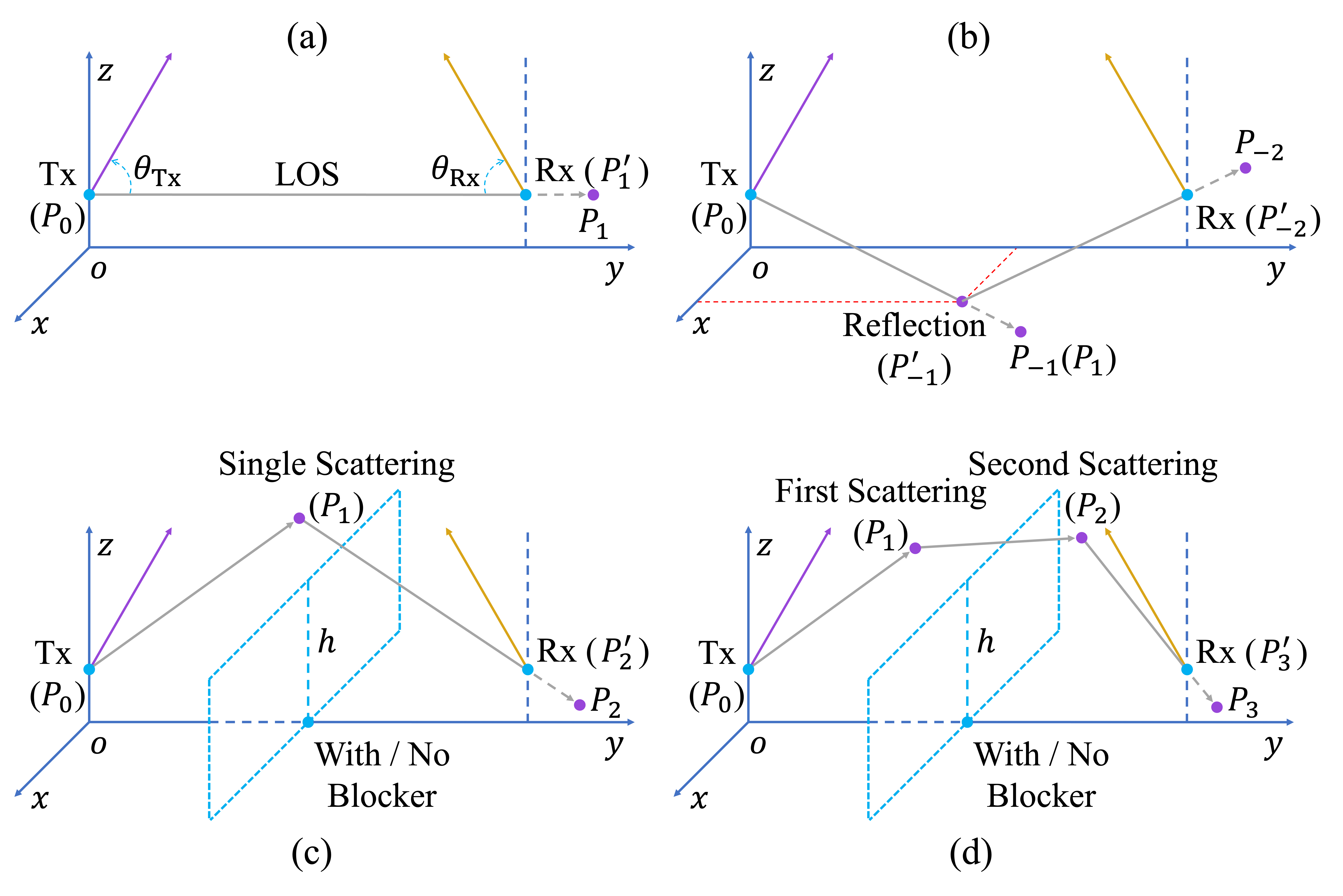}}
    \caption{Photon propagation models: (a) LOS path. (b) Ground reflection path. (c) Single-scattering path. (d) Multiple-scattering path.}
    \label{figure_mg}%
\end{figure}

In this section, a Monte Carlo (MC) simulation framework is developed to reproduce the experimental process and evaluate its agreement with the measured results. Figure~\ref{figure_mg} illustrates several photon-propagation geometries considered in the simulation. The XoY plane is defined as the ground surface. When LOS blockage is introduced, the obstacle is modelled as an infinitely wide vertical plane with height $h=3$~m located at $y=50$~m. The Tx and Rx are positioned at $(0~\mathrm{m},0~\mathrm{m},1~\mathrm{m})$ and $(0~\mathrm{m},100~\mathrm{m},1~\mathrm{m})$, respectively. Both Tx and Rx pointing directions lie within the YoZ plane without azimuthal misalignment, corresponding to the co-planar geometry used in the experiments. The Tx/Rx elevation angles are still denoted by $\theta_{\mathrm{Tx}}$ and $\theta_{\mathrm{Rx}}$. 

Figure~\ref{figure_mg}(a) illustrates a direct LOS propagation path under the no-blockage condition. A photon is emitted from point $P_{0}$ at the Tx. Although its first virtual scattering point is located at $P_{1}$, the photon reaches the Rx plane at $P_{1}'$ and is detected before the first scattering event occurs. Figure~\ref{figure_mg}(b) shows a single ground-reflection path, which also exists only without LOS blockage. The first virtual scattering point $P_{-1}$ is located below the ground plane, causing the photon to reflect at $P_{-1}'$. The photon then propagates toward the second virtual scattering point $P_{-2}$, but reaches the Rx plane at $P_{-2}'$ before the next scattering event occurs.

Figure~\ref{figure_mg}(c) illustrates a single-scattering propagation path, which can occur both with and without LOS blockage. The photon first undergoes a real scattering event at $P_{1}$ and is subsequently detected at $P_{2}'$ on the Rx plane before reaching the next virtual scattering point $P_{2}$. Similarly, Fig.~\ref{figure_mg}(d) shows a double-scattering path, where the photon is scattered at $P_{1}$ and $P_{2}$ before being detected at $P_{3}'$. In the simulation, photon reception after up to three scattering events is considered, although the triple scattering case is omitted from the figure for clarity. Higher-order scattering contributions were found to be negligible and were therefore omitted. In summary, direct LOS and single ground-reflection paths exist only under no-blockage conditions, whereas scattering-mediated propagation paths remain present in both blocked and unblocked scenarios.

\begin{figure}[!htbp]
    \centerline{\includegraphics[width=0.505\textwidth, keepaspectratio=true]{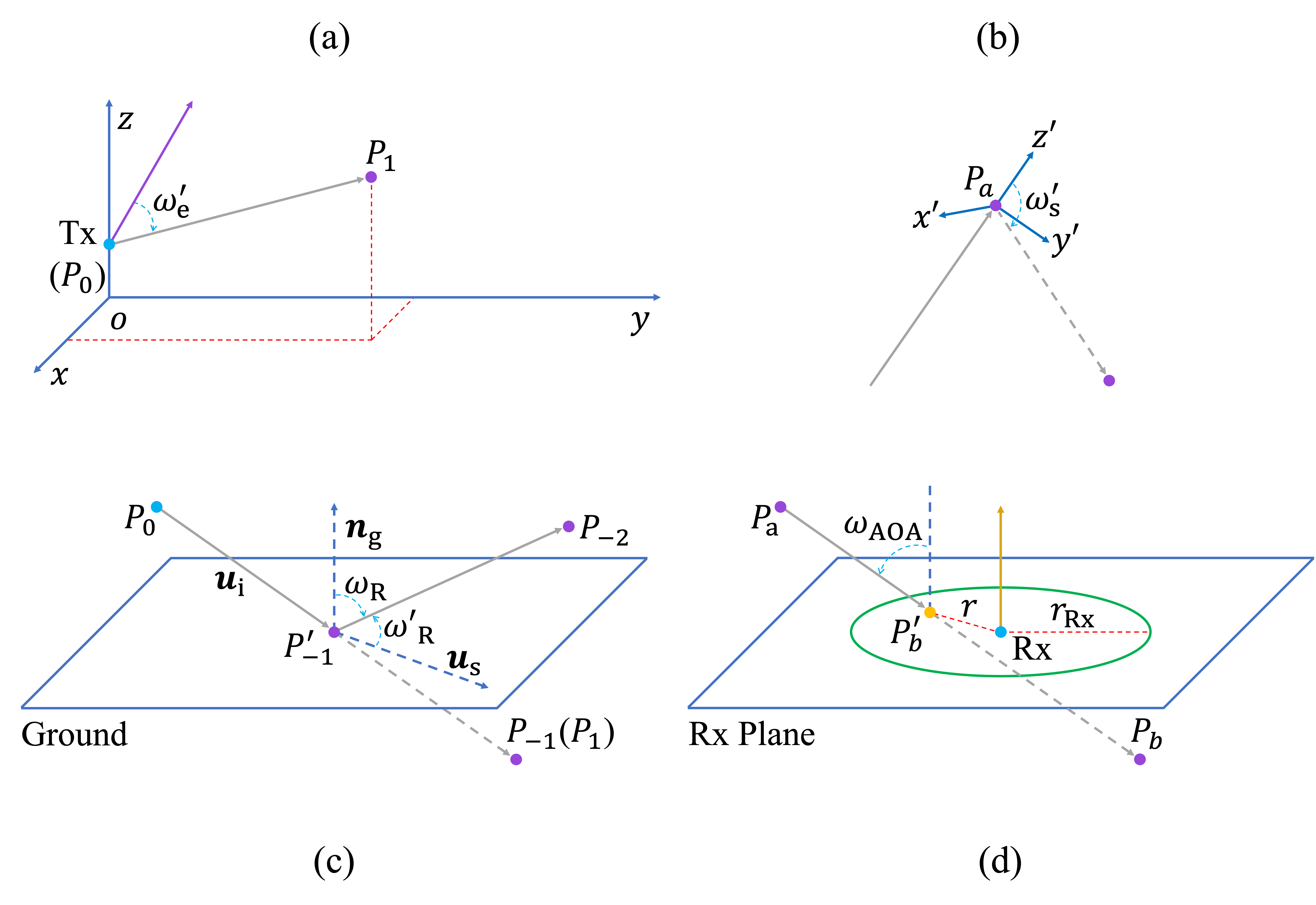}}
    \caption{Photon propagation algorithms: (a) Emission. (b) Scattering. (c) Ground reflection. (d) Reception.}
    \label{figure_pa}%
\end{figure}

Figure~\ref{figure_pa} illustrates the key computational procedures for photon emission, scattering, ground reflection, and reception in the proposed MC simulation framework.

As shown in Fig.~\ref{figure_pa}(a), a photon is emitted from point $P_{0}$ at the Tx. The angle between the photon propagation direction $\overrightarrow{P_{0}P_{1}}$ and the Tx pointing direction is defined as the local photon emission inclination angle $\omega_{\mathrm{e}}'$. Instead of assuming an idealised hard-boundary beam cone, the measured Tx irradiance profile $I_{\mathrm{Tx}}(\omega_{\mathrm{e}}')$ shown in Fig.~\ref{figure_sbd} is directly used to generate the photon emission distribution. The measured irradiance profile is first normalised to unit area to form the probability density function (PDF), 
\begin{equation}
p_{\mathrm{e}}(\omega_{\mathrm{e}}')
=
\frac{
I_{\mathrm{Tx}}(\omega_{\mathrm{e}}')
}
{
\int_{0}^{90}
I_{\mathrm{Tx}}(\xi)\,d\xi
},
\qquad
0^{\circ}\leq \omega_{\mathrm{e}}' \leq 90^{\circ}.
\label{eq:emission_pdf}
\end{equation}

The corresponding cumulative distribution function (CDF) is obtained as
\begin{equation}
F_{\mathrm{e}}(\omega_{\mathrm{e}}')
=
\int_{0}^{\omega_{\mathrm{e}}'}
p_{\mathrm{e}}(\xi)\,d\xi .
\label{eq:emission_cdf}
\end{equation}

The emission inclination angle is then generated using inverse-transform sampling:
\begin{equation}
\omega_{\mathrm{e}}'
=
F_{\mathrm{e}}^{-1}(U),
\qquad
U\sim\mathcal{U}(0,1).
\label{eq:emission_inverse_sampling}
\end{equation}

This procedure ensures that generated emission angles follow the experimentally measured Tx radiation pattern rather than an idealised beam-cone approximation. Since Tx radiation is azimuthally symmetric around the pointing direction, the local photon emission azimuth angle is uniformly distributed:
\begin{equation}
\varphi_{\mathrm{e}}'
=
360^{\circ}\times U,
\qquad
U\sim\mathcal{U}(0,1).
\label{eq:emission_azimuth}
\end{equation}

The local emission direction, represented by the polar angles $(\omega_{\mathrm{e}}',\varphi_{\mathrm{e}}')$, is subsequently transformed back into the global Cartesian coordinate system using direction cosines, yielding the photon propagation direction vector $\overrightarrow{P_{0}P_{1}}$ \cite{ding2009modeling}.

Figure~\ref{figure_pa}(b) shows the calculation of the photon propagation direction after a scattering event \cite{ding2009modeling}. The algorithm is similar to the emission-direction generation described above, except that the local photon scattering inclination angle $\omega_{\mathrm{s}}'$ is determined by the combined Rayleigh--Mie scattering distribution:
\begin{equation}
P_{\mathrm{combined}}(\omega_{\mathrm{s}}')
=
\frac{k_{\mathrm{s,Ray}}}{k_{\mathrm{s}}}
P_{\mathrm{Ray}}(\omega_{\mathrm{s}}')
+
\frac{k_{\mathrm{s,Mie}}}{k_{\mathrm{s}}}
P_{\mathrm{Mie}}(\omega_{\mathrm{s}}'),
\label{eq:combined_phase_function}
\end{equation}
where $k_{\mathrm{s}}=k_{\mathrm{s,Ray}}+k_{\mathrm{s,Mie}}$ is the total scattering coefficient.

The generalized Rayleigh scattering function is given by
\begin{equation}
P_{\mathrm{Ray}}(\omega_{\mathrm{s}}')
=
\frac{
3\left[
1+3\gamma+
(1-\gamma)\cos^{2}\omega_{\mathrm{s}}'
\right]
}
{16\pi(1+2\gamma)},
\label{eq:rayleigh_phase_function}
\end{equation}
and the generalized Henyey--Greenstein function for Mie scattering is expressed as
\begin{equation}
\begin{split}
P_{\mathrm{Mie}}(\omega_{\mathrm{s}}')
=
\frac{1-g^{2}}{4\pi}
\Bigg[
&\frac{1}
{\left(1+g^{2}-2g\cos\omega_{\mathrm{s}}'\right)^{3/2}} \\
&+
f\frac{3\cos^{2}\omega_{\mathrm{s}}'-1}
{2(1+g^{2})^{3/2}}
\Bigg].
\end{split}
\label{eq:mie_phase_function}
\end{equation}

Here, the model parameters are set to the commonly adopted values $\gamma=0.017$, $g=0.72$, and $f=0.5$ \cite{ding2009modeling}. The corresponding PDF is therefore obtained by direct normalisation:
\begin{equation}
p_{\mathrm{s}}(\omega_{\mathrm{s}}')
=
\frac{
P_{\mathrm{combined}}(\omega_{\mathrm{s}}')
}
{
\int_{0}^{180}
P_{\mathrm{combined}}(\xi)\,d\xi
},
\qquad
0^{\circ}\leq \omega_{\mathrm{s}}' \leq 180^{\circ}.
\label{eq:scattering_pdf}
\end{equation}

The remaining steps follow the same procedure as photon emission, including CDF construction, inverse-transform sampling, uniformly distributed azimuth generation, and direction-cosine rotation; these steps are omitted here for brevity.

Next, Fig.~\ref{figure_pa}(c) illustrates the ground-reflection process, which is modelled using the combined diffuse-specular Phong reflection model in \cite{cao2021reflection}. For the asphalt surface considered in this work, 95\% of incident photons are assumed to be absorbed and the remaining 5\% are reflected. The reflected field is represented by a weighted combination of Lambertian diffuse reflection and a directional specular Phong lobe, with the reflection direction $\overrightarrow{P_{-1}'P_{-2}}$ generated using the Monte Carlo sampling procedure described in \cite{cao2021reflection}. In this work, $\alpha=0.8$ and $m_{\mathrm{s}}=6$ are adopted, corresponding to a predominantly diffuse asphalt surface with a weak specular component.

Finally, Fig.~\ref{figure_pa}(d) illustrates the photon-reception process. The intersection point between the photon propagation path $\overrightarrow{P_{a}P_{b}}$ and the Rx plane is denoted as $P_{b}'$. The distance between $P_{b}'$ and the Rx centre is calculated as $r$. Assuming a circular receiver sensing window with radius $r_{\mathrm{Rx}}$, the photon is considered to enter the PMT if $r < r_{\mathrm{Rx}}$. However, the photon must first pass through the optical bandpass filter before entering the PMT. To model this process, the normalised Rx responsivity curve shown in Fig.~\ref{figure_sbd} is multiplied by the filter transmittance of 12\% at $0^{\circ}$ angle-of-arrival (AOA), forming the overall photon reception probability curve $p_{\mathrm{r}}(\omega_{\mathrm{AOA}})$ as a function of the incident AOA. The AOA $\omega_{\mathrm{AOA}}$ of the incident photon is first calculated and then the corresponding reception probability $p_{\mathrm{r}}(\omega_{\mathrm{AOA}})$ is obtained from the probability curve. Finally, a uniformly distributed random number $U\sim\mathcal{U}(0,1)$ is generated. The photon is considered successfully received by the PMT if $U < p_{\mathrm{r}}(\omega_{\mathrm{AOA}})$.

The remaining two parameters governing photon propagation, namely the scattering-event distance $d_{\mathrm{s}}$ and absorption distance $d_{\mathrm{a}}$, are generated as
\begin{equation}
d_{\mathrm{s}}
=
-\frac{\ln U_{1}}{k_{\mathrm{s}}},
\qquad
d_{\mathrm{a}}
=
-\frac{\ln U_{2}}{k_{\mathrm{a}}},
\label{eq:scattering_absorption_distance}
\end{equation}
where $U_{1},U_{2}\sim\mathcal{U}(0,1)$ are uniformly distributed random numbers, $k_{\mathrm{s}}$ and $k_{\mathrm{a}}$ denote the total atmospheric scattering coefficient and atmospheric absorption coefficient, respectively.

In addition to the experimental parameters listed in Table~\ref{table_kpoes}, the other key parameters used in the MC simulation are summarised in Table~\ref{table_kpomcs}. The atmospheric and ground-reflection parameters are selected based on \cite{ding2009modeling, cao2021reflection}.

\begin{table}[!htbp]
\centering
\caption{\label{table_kpomcs} Key Parameters of Monte Carlo Simulation.\\}
\begin{tabular}{|c|c|}
\hline
\textbf{Parameter} & \textbf{Value} \\
\hline
\hline
Total Number of Emitted Photons $N _{\textup{e}}$ & $5\times 10^{10} \sim 1\times 10^{12}$ \\
\hline
Atmospheric Absorption Coefficient $k_{\textup{a}}$ & $0.71~\textup{km}^{-1} $ \\
\hline
Rayleigh Scattering Coefficient $k_{\textup{s,Ray}}$ & $0.25~\textup{km}^{-1} $ \\
\hline
Mie Scattering Coefficient $k_{\textup{s,Mie}}$ & $0.28~\textup{km}^{-1} $ \\
\hline
Phase Function Parameters $\gamma$ / $g$ / $f$ & $0.017$ / $0.72$ / $0.5$ \\
\hline
Ground Reflectance $\rho_{\textup{g}}$ & $5\%$ \\
\hline
Phong Model Parameters $\alpha$ / $m_{\textup{s}}$ & $0.8$ / $6$ \\
\hline
\end{tabular}
\end{table}

\begin{figure*}[!htbp]
    \centering
    \includegraphics[width=1.02\textwidth]{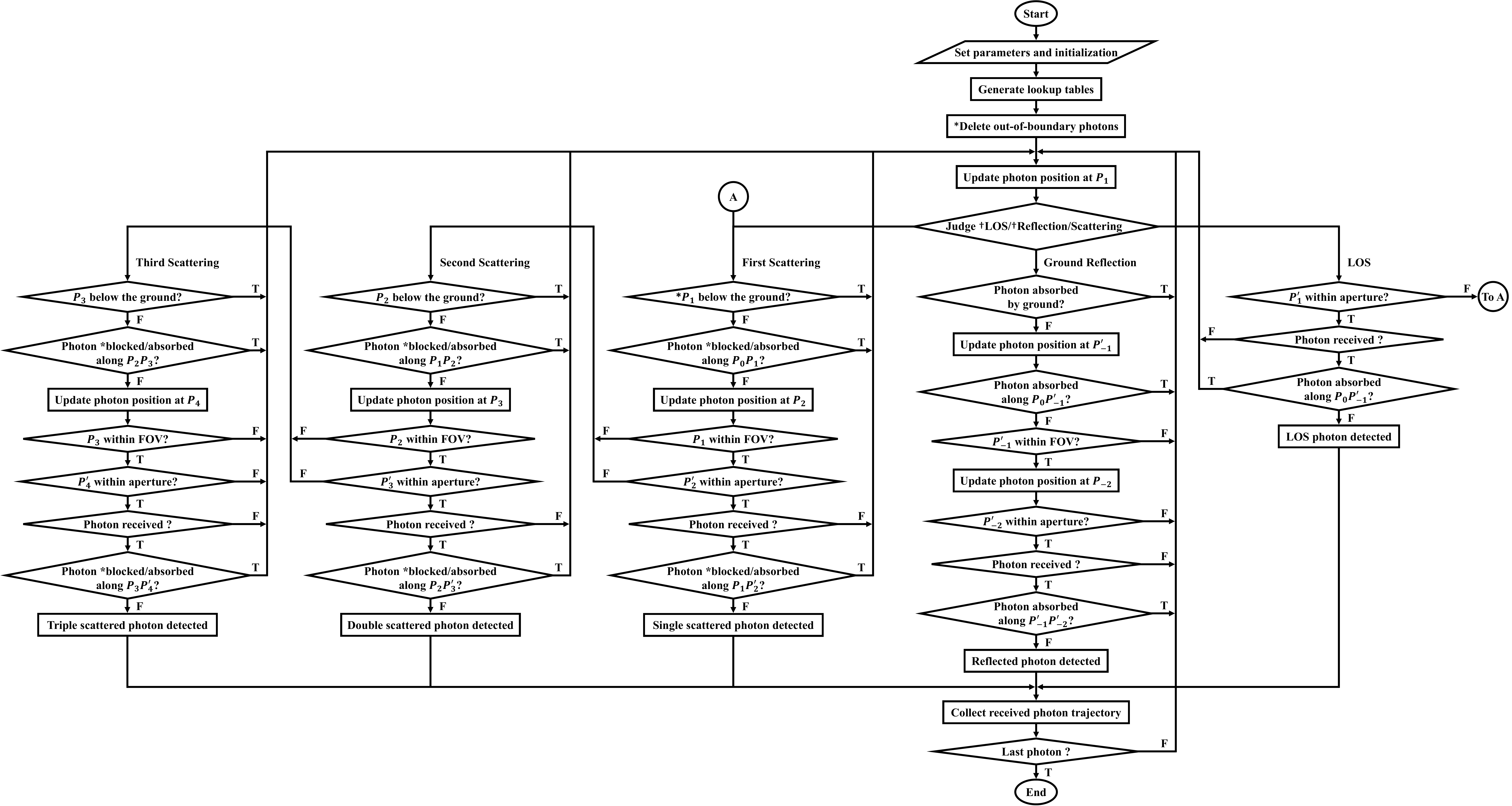}
    \caption{Logic chart of direct-photon-tracing MC simulation. (*: with obstacle. $\dagger$: no obstacle.)}
    \label{figure_lcmc}
\end{figure*}

After introducing the propagation algorithms, the overall workflow of the MC simulation is illustrated in Fig.~\ref{figure_lcmc}. Following parameter initialisation, several look-up tables (LUTs) are pre-generated for important random variables, including $(\omega_{\mathrm{e}}',\varphi_{\mathrm{e}}')$, $(\omega_{\mathrm{s}}',\varphi_{\mathrm{s}}')$, $d_{\mathrm{s}}$, and $d_{\mathrm{a}}$. This avoids repeatedly calling random-number generators during photon tracing and significantly improves the simulation efficiency.

Next, a maximum first-scattering distance limit $d_{\mathrm{s},P_{0}P_{1},\mathrm{max}}$ is introduced. If a photon propagates beyond this distance without scattering, it is assumed unlikely to be received in the scattering-mediated NLOS link and is therefore discarded. This removes out-of-boundary photons and substantially reduces the LUT memory requirement.

The location of the first scattering point $P_{1}$, together with the presence or absence of an LOS blocker, is then used to determine whether the photon enters the LOS, ground-reflection, or first-scattering branch. The detailed logic of each branch is shown in Fig.~\ref{figure_lcmc} and is not repeated here for clarity. Note that the steps marked with ``$*$'' are omitted when no LOS blocker is present. Once a photon is successfully received, all information associated with its propagation path is recorded before tracing the next photon. The simulation continues until all photons have been processed.

From the photon-reception procedure in Fig.~\ref{figure_pa}(d) and the overall logic chart, it can be seen that the proposed photon-tracing MC simulation adopts a direct-reception approach. In this method, each photon is explicitly traced and directly judged for reception by Rx, rather than using the reception-probability approach commonly adopted in the literature \cite{ding2009modeling, drost2011uv}, where only the probability of photon reception is evaluated at each propagation step. Although the direct-reception approach is computationally less efficient than probability-based reception methods due to the small Rx sensing area and consequently low photon-hit rate, it provides several advantages. Since the complete propagation history of each received photon is explicitly recorded, the method offers greater flexibility for trajectory-resolved analysis, three-dimensional (3D) visualisation of photon propagation paths, and more detailed investigation of different propagation mechanisms. To compensate for the increased computation time, the simulations were executed on three remote multi-core CPU servers in parallel. Depending on the Tx/Rx geometry, each simulation point typically required approximately half a day to two days to achieve satisfactory convergence, which remained computationally manageable for this work.

\subsection{Total Path Loss and Detected Photon Components}\label{subsection_tpladpc}

\begin{figure}[!htbp]
    \centerline{\includegraphics[width=0.51\textwidth, keepaspectratio=true]{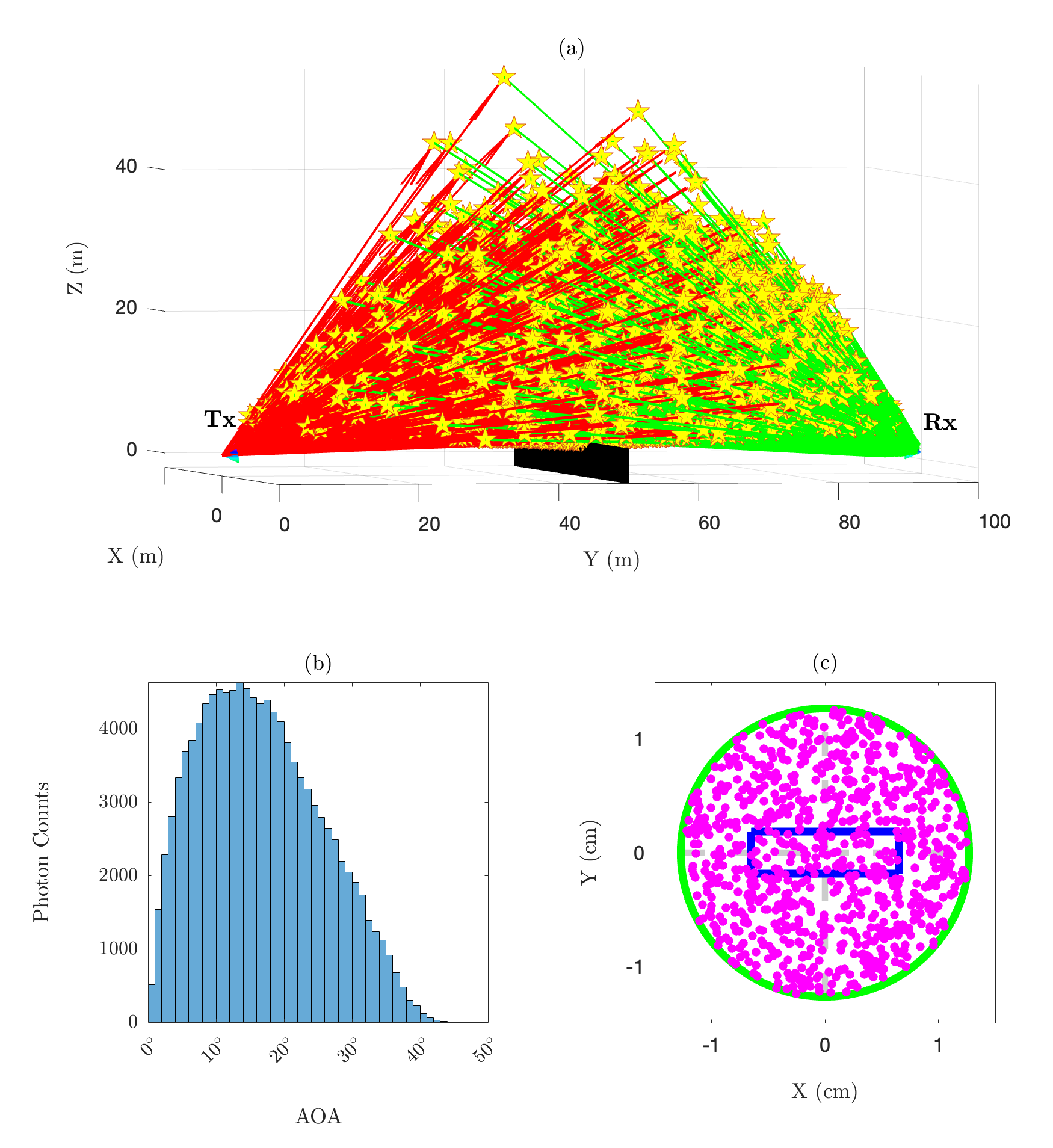}}
    \caption{Monte Carlo simulation results for each link geometry: (a) Photon trajectories of single-scattered photons for Tx$_{15^{\circ}}$/Rx$_{15^{\circ}}$. (b) AOA distribution for Tx$_{45^{\circ}}$/Rx$_{45^{\circ}}$. (c) Received single-scattered photon positions on Rx plane for Tx$_{45^{\circ}}$/Rx$_{45^{\circ}}$. (Green circle: Simulated sensing area. Blue rectangle: PMT sensing window.)} 
    \label{figure_pt}%
\end{figure}

For each Tx/Rx elevation-angle combination, the MC simulation provides the photon trajectory information, total path loss, photon-component contributions from different propagation mechanisms, and the channel impulse response (CIR). Figure~\ref{figure_pt}(a) shows the trajectories of single-scattered photons for the Tx$_{15^{\circ}}$/Rx$_{15^{\circ}}$ geometry. It can be clearly observed that all scattered photons correctly propagate above the obstacle before reaching Rx. The scattering points form an irregular 3D density cloud distributed throughout the propagation region. And the significant path-length differences among numerous scattering trajectories give rise to severe temporal dispersion and multipath broadening in NLOS UVC channels.

Figure~\ref{figure_pt}(b) presents the AOA distribution of all received photons for Tx$_{45^{\circ}}$/Rx$_{45^{\circ}}$. Interestingly, relatively few photons arrive exactly along the Rx pointing direction (at $0^{\circ}$ AOA), while the dominant AOA peak appears around $15^{\circ}$. However, commercially available optical bandpass filters based on multilayer thin-film interference typically exhibit significantly reduced transmittance at larger AOAs, which decreases the number of UV photons that can pass through the filter. This observation suggests that future optical-filter development for NLOS UVC systems may benefit from wider-FOV and higher-transmittance designs. Figure~\ref{figure_pt}(c) shows the spatial distribution of received photons on the Rx plane. In the MC simulation, the green Rx sensing area $A_{\mathrm{sim}}$ is modelled as a circular aperture with radius $r_{\mathrm{Rx}}$, whereas the actual blue PMT sensing area $A_{\mathrm{PMT}}$ is a rectangular window of $0.37\times1.3~\mathrm{cm}$. Their area ratio is defined as $ k = \frac{A_{\mathrm{sim}}}{A_{\mathrm{PMT}}}$. Accordingly, the total path loss of simulated system in dB is calculated as
\begin{equation}
\begin{split}
L_{\mathrm{total}}
&=L_{\mathrm{channel}}+L_{\mathrm{Tx}}+L_{\mathrm{Rx}} \\
&=10\log_{10}\left(\frac{kN_{\mathrm{emit}}}{\eta_{\mathrm{reflector}}\eta_{\mathrm{PMT}}N_{\mathrm{received}}}\right).
\end{split}
\label{eq:mc_total_path_loss}
\end{equation}
where $N_{\mathrm{emit}}$ and $N_{\mathrm{received}}$ are the emitted and received photon numbers. The reflector efficiency $\eta_{\mathrm{reflector}}$ and PMT detection efficiency $\eta_{\mathrm{PMT}}$ are listed in Table~\ref{table_kpoes}, while the optical bandpass-filter transmittance has already been incorporated into the photon-reception process during simulation.

It can also be observed from Fig.~\ref{figure_pt}(c) that the received photons are approximately uniformly distributed within a certain region on the Rx plane. Therefore, during simulation, the value of $r_{\mathrm{Rx}}$ can be intentionally increased slightly to enlarge the effective sensing area, allowing more photons to be collected without increasing the emitted photon number $N_{\mathrm{emit}}$. This significantly improves the convergence speed of result while maintaining good accuracy. In addition, the resulting CIR is largely insensitive to moderate adjustments of $r_{\mathrm{Rx}}$, whereas the larger number of received photons further improves the smoothness of the simulated CIR curves. This serves as another effective approach for improving the computational efficiency of the direct-reception MC simulation.

\begin{figure}[!htbp]
    \centerline{\includegraphics[width=0.5\textwidth, keepaspectratio=true]{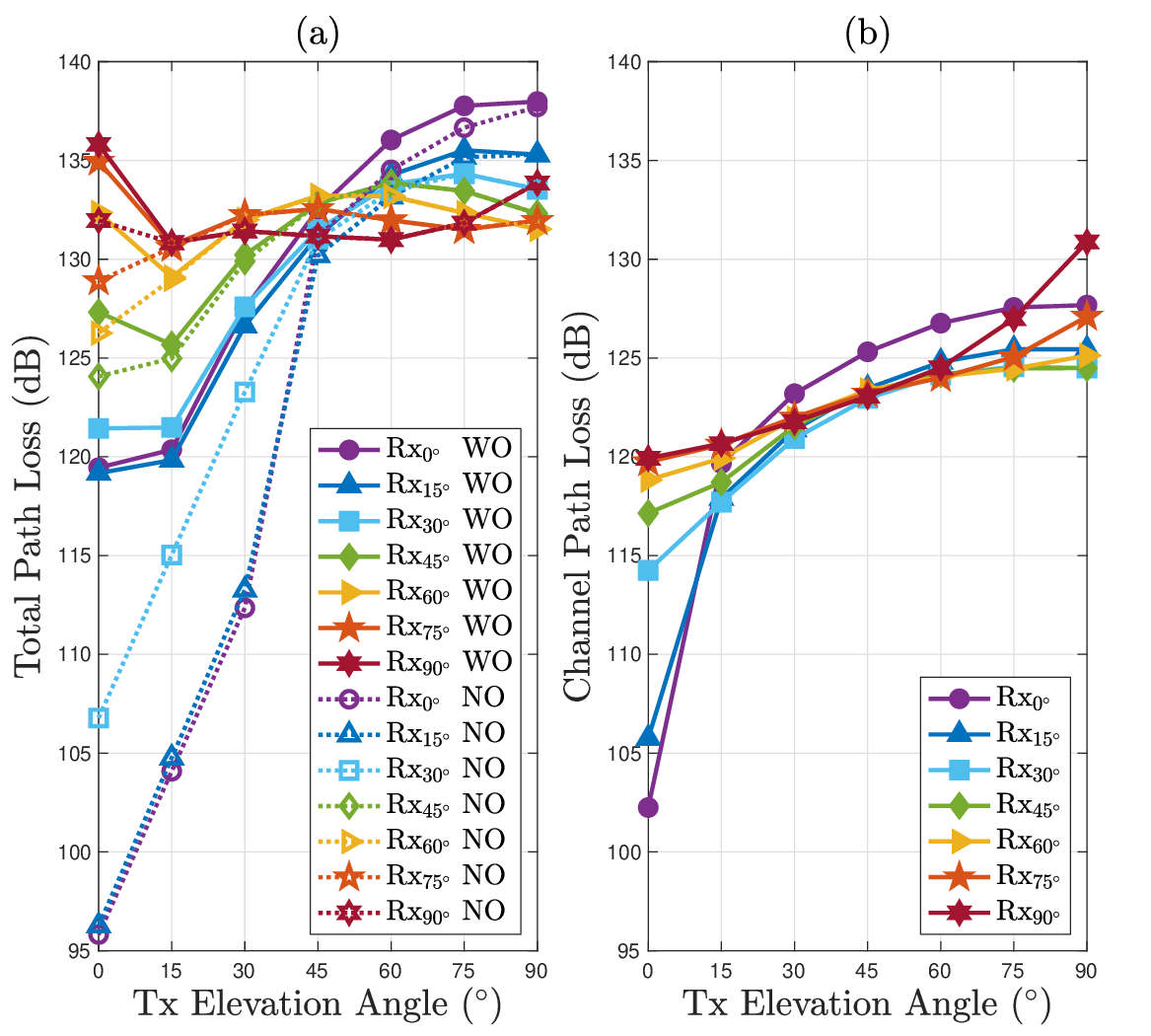}}
    \caption{(a) Total path loss from MC simulation. (b) Channel path loss from analytical modeling. (WO: with obstacle / NO: no obstacle)}
    \label{figure_tplsa}%
\end{figure}

The total path loss results obtained from the MC simulation are presented in Fig.~\ref{figure_tplsa}(a) and Fig.~\ref{figure_tples}(b). Compared with the experimentally measured results shown in Fig.~\ref{figure_tples}(a), both the absolute values and overall trends exhibit very good agreement. Since the atmospheric-scattering and ground-reflection parameters adopted in the simulation are based on commonly used values from the literature \cite{ding2009modeling, cao2021reflection}, whereas the actual scattering, absorption, and reflection properties of the measurement environment are not precisely known, some discrepancy between the simulation and experimental results is inevitable.

For comparison, the classical analytical model reported in \cite{luettgen1991non} was also reproduced with the same experimental parameters. This model represents one of the earliest and most widely referenced analytical formulations for scattering channels and is commonly used as a benchmark in the literature. The predicted channel path loss $L_{\mathrm{channel}}$ is shown in Fig.~\ref{figure_tplsa}(b). It can be observed that this analytical model captures the general trend of the path-loss variation reasonably well, whereas the proposed MC simulation provides substantially improved agreement in the detailed behaviour. The discrepancy of the simplified analytical model mainly arises from the accumulation of multiple simplifying assumptions, including idealised beam/FOV cones, the absence of obstacle blockage, and the single-scattering approximation. In particular, direct LOS propagation can still exist in low-elevation geometries such as Tx$_{0^{\circ}}$/Rx$_{0^{\circ}}$ and Tx$_{0^{\circ}}$/Rx$_{15^{\circ}}$, leading to noticeably underestimated path-loss predictions.

To quantitatively evaluate the agreement between simulation and experiment, the root-mean-square error (RMSE) and correlation coefficient, $\rho$, were calculated using all measured and simulated path-loss values across the complete set of Tx/Rx elevation-angle combinations. For fair comparison, an additional 16.9~dB Tx/Rx hardware loss, originating from the reflector efficiency, optical filter transmittance, and PMT detection efficiency, is added to the channel path loss predicted by the analytical model to obtain the corresponding system total path loss. The proposed MC simulation achieves an RMSE of only 2.9~dB with a correlation coefficient of $\rho=0.89$, whereas the analytical model gives a substantially larger RMSE of 7.0~dB with a lower correlation coefficient of $\rho=0.78$. These results demonstrate that the proposed MC simulation framework provides significantly improved agreement with the experimental measurements in terms of both absolute path-loss prediction and overall trend reproduction.

\begin{figure}[!htbp]
    \centerline{\includegraphics[width=0.53\textwidth, keepaspectratio=true]{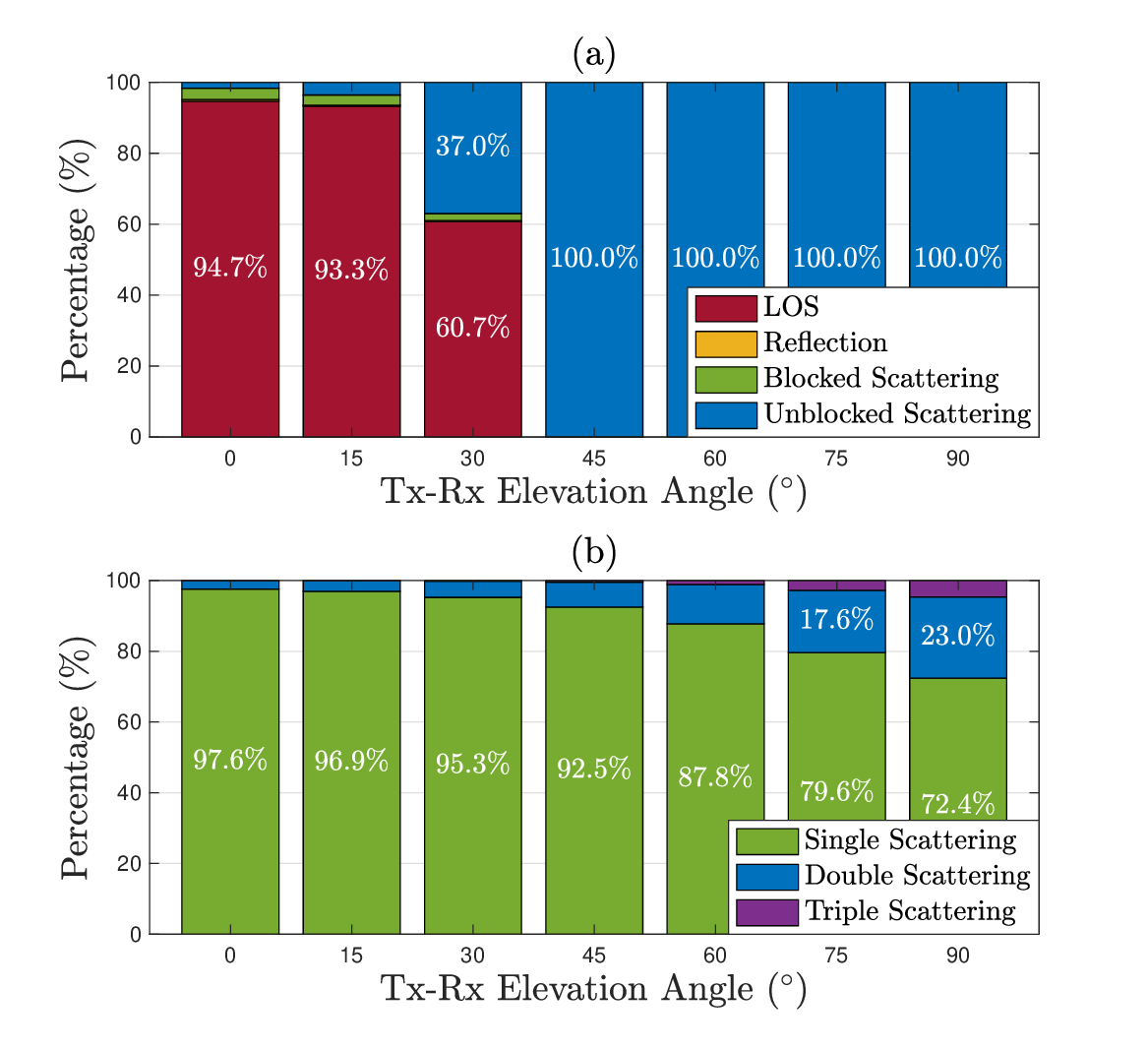}}
    \caption{Detected photon components for varying Tx-Rx elevation angles: (a) No LOS blocker. (b) With LOS blocker.}
    \label{figure_dpc}%
\end{figure}

Next, the composition of the received photons is further analysed. Seven representative co-planar geometries are selected, where $\theta_{\mathrm{Tx}}=\theta_{\mathrm{Rx}}$ jointly increases from $0^{\circ}$ to $90^{\circ}$. The four photon components under the no-blockage condition, namely LOS photons, ground-reflection photons, near-ground scattering photons that would be blocked by the obstacle, and scattering photons that can bypass the obstacle, are summarised in Fig.~\ref{figure_dpc}(a).

For the Tx$_{0^{\circ}}$/Rx$_{0^{\circ}}$ geometry, LOS photons dominate the received signal and account for around 95\% of all received photons. Even at Tx$_{30^{\circ}}$/Rx$_{30^{\circ}}$, the LOS component still contributes 61\% of the total received photons. The LOS component only disappears when the elevation angles exceed approximately Tx$_{45^{\circ}}$/Rx$_{45^{\circ}}$. These results quantitatively demonstrate that, for field measurements employing diffused beams, the received signal can still contain a substantial amount of diffused-LOS photons unless explicit LOS blockage is introduced. Consequently, the measured results under such conditions may not fully represent purely scattering-mediated NLOS channel characteristics.

The contribution of ground reflection is comparatively weak. Although it reaches its maximum value at Tx$_{0^{\circ}}$/Rx$_{0^{\circ}}$, the contribution remains only 0.4\% and is therefore negligible in the present experimental setup. In contrast, the scattered-photon composition changes significantly with the Tx/Rx elevation angles. At Tx$_{0^{\circ}}$/Rx$_{0^{\circ}}$, approximately 66\% of scattered photons originate from near-ground small-angle scattering paths, which can also be blocked by the 3-m obstacle. As $\theta_{\mathrm{Tx}}$ and $\theta_{\mathrm{Rx}}$ increase beyond $45^{\circ}$, most scattered photons can successfully propagate above the obstacle, indicating a transition toward genuinely obstacle-bypassing ability.

Additionally, Fig.~\ref{figure_dpc}(b) further presents the composition of single- and multiple-scattered photons under LOS-blocked conditions. As $\theta_{\mathrm{Tx}}$ and $\theta_{\mathrm{Rx}}$ jointly increase from $0^{\circ}$ to $90^{\circ}$, the contribution of single-scattered photons decreases from 98\% to 72\%, although it remains the dominant component throughout all geometries. Meanwhile, the contribution of double-scattered photons increases from only 2.4\% at Tx$_{0^{\circ}}$/Rx$_{0^{\circ}}$ to 23\% at Tx$_{90^{\circ}}$/Rx$_{90^{\circ}}$. Triple-scattered photons remain the weakest component, but their contribution still increases from a negligible 0.04\% to 4.6\% at Tx$_{90^{\circ}}$/Rx$_{90^{\circ}}$. Overall, increasing the Tx/Rx elevation angles gradually enhances the contribution of higher-order scattering components. Nevertheless, single scattering remains the dominant propagation mechanism throughout the 100-m NLOS link investigated in this work.

\subsection{Channel Impulse Response and Bandwidth}\label{subsection_cirab}

From the total propagation distance of each received photon, the corresponding propagation time can be calculated. By accumulating the received photon counts within discrete time bins (\textit{e.g.} 200~ps), the channel impulse response (CIR) of each link geometry can be directly reconstructed. In the following analysis, only LOS-blocked link geometries are considered.

\begin{figure}[!htbp]
    \centerline{\includegraphics[width=0.5\textwidth, keepaspectratio=true]{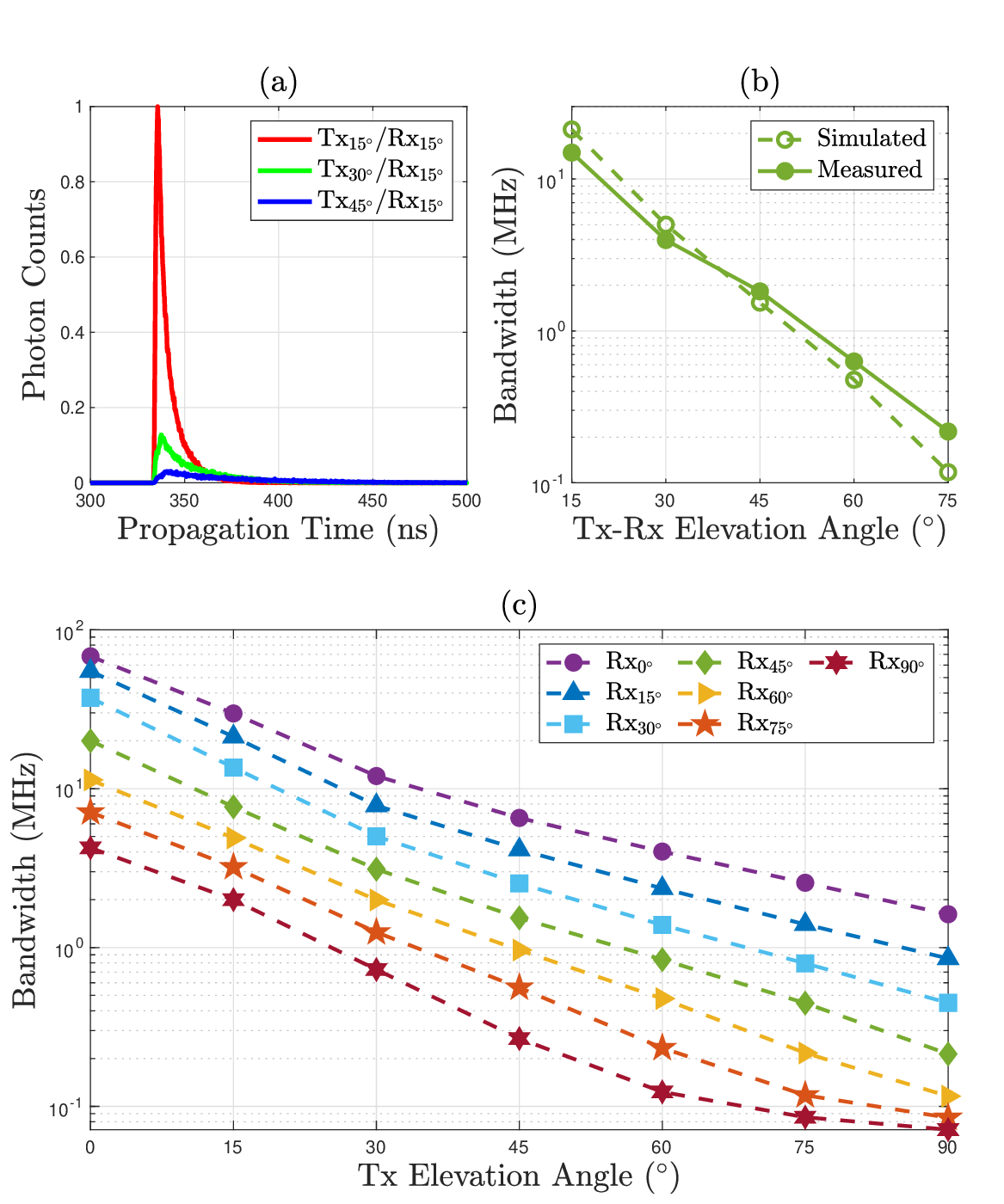}}
    \caption{CIRs and bandwidth of the 100-m LOS-blocked scattering channel. (a) Simulated CIRs. (b) Simulated and measured -3 dB bandwidth. (c) Simulated bandwidth for varying Tx/Rx elevation angles.}
    \label{figure_cirbw}%
\end{figure}

Figure~\ref{figure_cirbw}(a) presents the normalised CIRs for three different geometries. The Tx$_{15^{\circ}}$/Rx$_{15^{\circ}}$ exhibits the highest CIR peak and the smallest root-mean-square (RMS) delay spread $\tau_{\mathrm{rms}}$ of only 9.0~ns. As $\theta_{\mathrm{Tx}}$ increases to $30^{\circ}$ and $45^{\circ}$, the CIR peak decreases significantly, while the RMS delay spread broadens to 20.3~ns and 35.6~ns, respectively. Indicates both the rapid increase in path loss and the increasingly severe pulse-broadening effect at higher elevation angles.

The frequency response is subsequently obtained by applying a fast Fourier transform (FFT) to the CIR, from which the $-3$~dB bandwidth can be directly extracted. The simulated bandwidths for all 49 Tx/Rx elevation-angle combinations are given in Fig.~\ref{figure_cirbw}(c). The maximum bandwidth of 68.1~MHz is achieved at Tx$_{0^{\circ}}$/Rx$_{0^{\circ}}$, whereas it drops to only 71.4~kHz at Tx$_{90^{\circ}}$/Rx$_{90^{\circ}}$. The bandwidth variation across different link geometries spans nearly three orders of magnitude. In addition, the bandwidth decreases continuously as either the Tx or Rx elevation angle increases, indicating increasingly severe multi-path effects of the scattering channel.

Furthermore, the measured bandwidths for the five representative geometries are compared with the MC-simulated results in Fig.~\ref{figure_cirbw}(b). Excellent agreement is observed, further validating the accuracy and reliability of both the proposed measurement method and the MC simulation framework.

%\subsection{Influence of Link Distance}\label{subsection_iotrsd}

%\begin{figure}[!htbp]
%    \centerline{\includegraphics[width=0.5\textwidth, keepaspectratio=true]{Figures/distance.eps}}
%    \caption{(a) Total path loss vs. Tx-Rx separation distance. (b) Channel bandwidth vs. Tx-Rx separation distance.}
%    \label{figure_dis}%
%\end{figure}%

% ========== Data Transmission ====================================
\section{Experimental LOS-Blocked NLOS UVC Data Transmission at 100 m}\label{section_data_transmission}

As demonstrated in the preceding experimental and simulation studies, LOS-blocked scattering-mediated NLOS UVC links are characterised by extremely high path loss and limited channel bandwidth, both of which constrain the achievable data rate. To compensate for the high path loss, the emitted optical power was doubled to 1.8~W by duplicating the LED array, while a 48-cm-diameter parabolic reflector was added in front of the PMT to increase its effective photon-collection area. In addition, the original optical bandpass filter used for path loss measurement, which had only 12\% transmittance at 265~nm, was replaced by a higher-transmittance bandpass filter (Edmund \#39-338) with $>$50\% transmittance, resulting in more than a fourfold increase in the received photon count.

\begin{figure}[!htbp]
    \centerline{\includegraphics[width=0.5\textwidth, keepaspectratio=true]{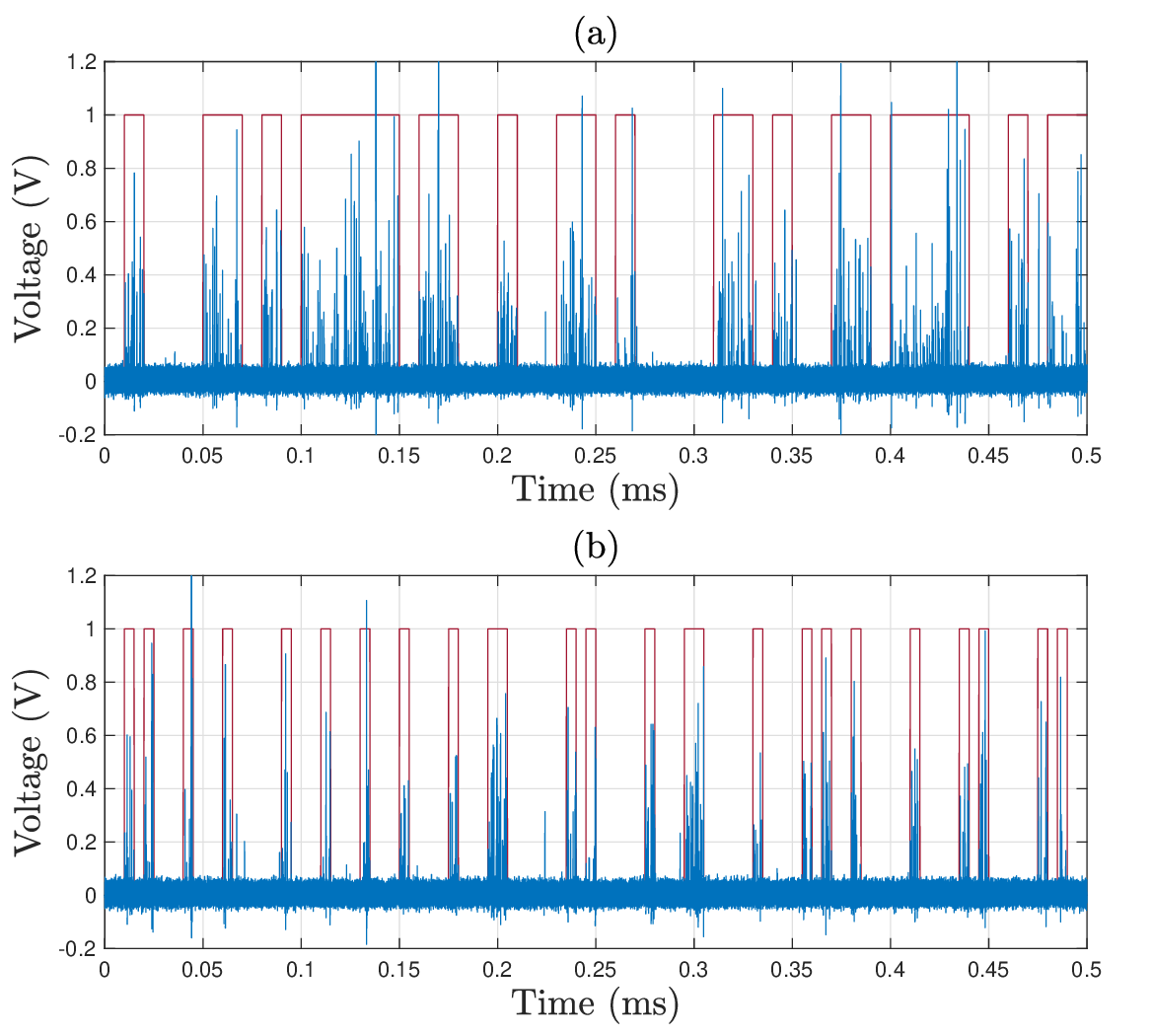}}
    \caption{Typical PMT output signals consisting of discrete photon-current spikes and the corresponding reference symbols. (a) OOK-NRZ. (b) 4-PPM. (Blue: PMT outputs. Red: reference symbols.)}
    \label{figure_pmtos}%
\end{figure}

The data transmission experiments were conducted in the late afternoon of 06/04/2026 under clear-sky conditions. The atmospheric condition was similarly clean to that observed on 04/03/2026 and the background PCR was 5360~counts/s. The link geometry remained unchanged, with a 100~m co-planar link distance and identical $\theta_{\mathrm{Tx}}$ and $\theta_{\mathrm{Rx}}$ jointly varying from $0^{\circ}$ to $90^{\circ}$. Two modulation schemes, namely OOK-NRZ and 4-PPM, were experimentally evaluated using photon-counting detection at four bit rates of 20, 50, 100, and 200~kbps. Typical sections of the received PMT output signal, consisting of discrete photon-current spikes together with the synchronized reference symbols, are shown in Fig.~\ref{figure_pmtos}(a) and Fig.~\ref{figure_pmtos}(b).

\begin{figure}[!htbp]
    \centerline{\includegraphics[width=0.52\textwidth, keepaspectratio=true]{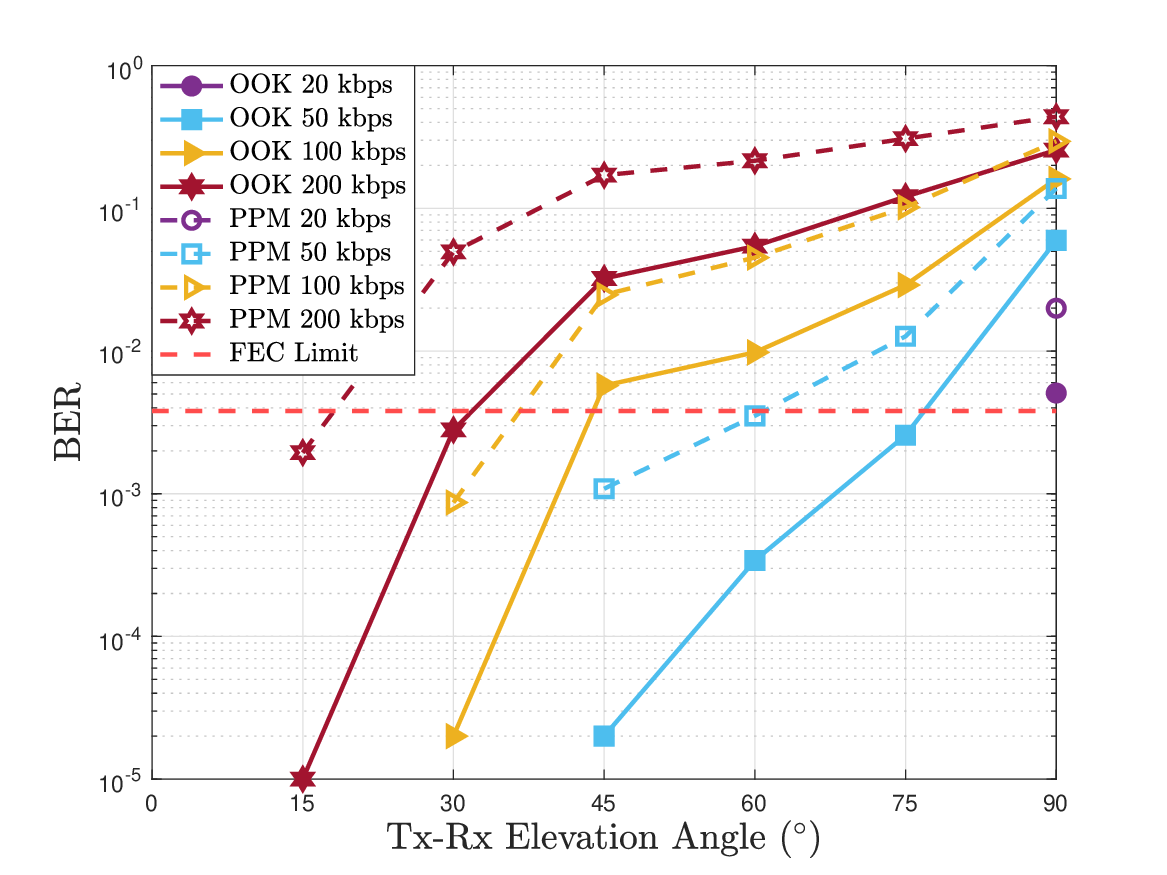}}
    \caption{BER performance of the 100-m LOS-blocked scattering NLOS UVC link under varying Tx/Rx elevation angles. (a) OOK-NRZ. (b) 4-PPM.}
    \label{figure_btreaudms}%
\end{figure}

The measured BER results are presented in Fig.~\ref{figure_btreaudms}. The red dashed line indicates the pre-forward error correction (FEC) threshold of $\mathrm{BER}=3.8\times10^{-3}$, below which the received signal is considered sufficiently reliable for further error correction. Due to the limited number of transmitted bits ($1\times10^{5}$--$1\times10^{6}$), the BER measurement resolution was limited. Consequently, BER values below $1\times10^{-5}$ were treated as error-free within the measurement resolution.

For a fixed modulation scheme, higher Tx/Rx elevation angles increase the channel path loss, leading to degraded signal quality and higher BER values. As a result, the low-elevation-angle geometries can support reliable data transmission at 200~kbps and possibly above, whereas higher-elevation-angle geometries only support reliable transmission at approximately 20--50~kbps. Furthermore, comparing the two modulation methods, although the background noise level remained the same, the average emitting optical power of OOK-NRZ was twice that of 4-PPM. Consequently, more signal photons are received under OOK-NRZ transmission, resulting in slightly improved performance and lower BER compared with 4-PPM.

% ========== Conclusion====================================
\section{Conclusion} \label{section_conclusion}

This work presented a comprehensive experimental and numerical investigation of explicitly obstacle-blocked, scattering-mediated NLOS UVC system over a 100~m outdoor link. For the first time, a systematic comparison between LOS-blocked and open-field UVC links was conducted, demonstrating that noticeable diffused-LOS components can remain present in practical wide-beam UVC systems. The results further showed that conventional FWHM-based BDA/FOV criteria do not necessarily guarantee strictly scattering-mediated propagation. A photon-tracing Monte Carlo framework incorporating realistic hardware characteristics and obstacle geometry achieved excellent agreement with the measured path loss and channel bandwidth under realistic propagation conditions. Despite the extremely high path loss and severe bandwidth limitation of strictly LOS-blocked scattering-mediated channels, reliable 100-m transmission at data rates of 20--200~kbps was successfully demonstrated using photon-counting detection under varying Tx/Rx elevation angles. To the best of the authors' knowledge, this represents the longest transmission distance and highest demonstrated data rate reported for explicitly obstacle-blocked scattering-mediated NLOS UVC links. These results provide important insights into the practical performance limits of scattering-mediated UVC systems and establish rigorous experimental benchmarks for future solar-blind OWC research.

% ========== references section ====================================
\bibliographystyle{IEEEtran}
\bibliography{References}

\end{document}